\documentclass[a4paper,fleqn,usenatbib,useAMS]{mnras}

\usepackage[T1]{fontenc}
\usepackage{ae,aecompl}

\usepackage{graphicx}
\usepackage{amsmath}	
\usepackage{amssymb}	
\usepackage{multicol}
\usepackage{xcolor}
\usepackage{pdflscape}	
\usepackage{hyperref}
\usepackage{ulem}
\usepackage{soul}

\title[]{Hot graphite dust in the inner regime of NGC 4151}

\author[Subhashree Swain et al.]{
Subhashree Swain$^{1}$,
P. Shalima$^{2}$,
K.V.P. Latha$^{1}$\thanks{E-mail: lathakvp@gmail.com}
and Krishna B. S. Swamy$^{3}$
\\
$^{1}$Pondicherry University, Puducherry, 605014,India\\
$^{2}$Manipal Centre for Natural Sciences, Centre of Excellence, Manipal Academy of Higher Education, Manipal, Karnataka 576104, India\\
$^{3}$Ahmedabad University,Ahmedabad 380009, India\\
}

\date{Accepted XXX. Received YYY; in original form ZZZ}
\pubyear{2019}

\begin{document}
\label{firstpage}
\pagerange{\pageref{firstpage}--\pageref{lastpage}}
\maketitle

\begin{abstract}
	We model the near infrared SED of NGC 4151 with a 3-D radiative
transfer SKIRT code, using which torus only (TO) and Ring And Torus (RAT) scenarios are studied. In the RAT models, a graphite ring-like structure (clumpy or smooth), is incorporated between the torus and the accretion disk.
We vary the inclination angle $(i)$, inner radius (of the torus and the ring, $R_{\rm in, t}$ and $R_{\rm in, r}$ respectively), torus half-opening angle ($\sigma $), optical depth ($\tau_{9.7, \rm t} $ of the torus and $\tau_{9.7, \rm r} $ of the ring ) and the dust clump size ($R_{\rm clump}$).
We perform a statistical analysis of the parameter space and find that all the models are able to explain the flat NIR SED of NGC 4151 with minor differences in the derived parameters. For the TO model, we get, $R_{\rm in, t}=0.1$ pc,  $\sigma = 30^\circ $, $i = 53^\circ $, $\tau_{9.7, \rm t}=10$ and the clumpsize, $R_{\rm clump}$ =0.4 pc. For the smooth RAT model, $R_{\rm in, \rm r}=0.04$ pc, $\tau_{9.7, \rm total}$ = 11 and for the clumpy RAT model, $R_{\rm in, r} = 0.04$ pc/0.06 pc and $\tau_{9.7, \rm total}=20$.
The $R_{\rm in, t}$ from the TO model does not agree with the NIR observations ($\sim 0.04$ pc).
Hence, the most likely scenario is that a hot graphite ring is located at a distance 0.04 pc from the centre, composed of a smooth distribution of dust 
	followed by a dusty torus at 0.1 pc with ISM type of grains.
\end{abstract}




\section{Introduction}
\label{sect:intro}
Active galaxies, unlike normal galaxies emit a tremendous amount of radiation throughout the electromagnetic spectrum with peak emission 
in the mid-infrared (MIR). This radiation has been observed to be non-stellar in origin. The nuclei of active galaxies are called active 
galactic nuclei (AGN) and are one of the brightest objects in the universe with luminosities $\backsim$ $10^{44}-10^{46}$ ergs\ s$^{-1}$ as 
they are powered by a supermassive blackhole ($\backsim$\,10$^{6}$-10$^{9}$\,M$_\odot$). According to the most widely accepted model of 
AGN, the central region of an AGN consists of an accretion disk that is surrounded by a geometrically and optically thick putative 
dusty structure, which is clumpy and toroidal, often referred to as the dusty torus. The torus is thought to be located at a radial 
distance of $0.1-10$ pc \citep{doi:10.1146/annurev-astro-082214-122302} from the central blackhole. The accretion disk mainly emits in 
the UV/optical regime (0.1$-$0.8 $\mu$m) and this radiation is absorbed by the dust grains in the torus. These heated dust grains then 
re-emit thermal radiation in the near-infrared (NIR) (1-5\, $\mu$m), MIR(5-25\, $\mu$m) and far-infrared (FIR) (25-380\, $\mu$m) regimes resulting in the observed 
IR spectral energy distribution (SED) of AGN. 

AGN are classified as type 1 and type 2 Seyfert nuclei on the basis of observed broad and narrow emission lines respectively.  The AGN unification scheme \citep{1985ApJ...297..621A, antonucci1993unified} suggests that the Type 1 and Type 2 AGN are a single entity with the same physics but are only different in their orientation with respect to the observer. In Type 2 AGN, the radiation from the accretion disk is blocked by the dusty material present in the torus, while in Type 1 AGN we directly observe the innermost regions of the AGN.

Therefore, by studying the IR SED of AGN one can get some insights into the dust morphology of AGN. The central dusty region termed as torus has been found to have sizes ranging from $0.1-10$\, pc from MIR imaging \citep{2005ApJ...618L..17P,2008ApJ...681..141R} and interferometry \citep{refId0}, where the innermost radius is determined by the dust sublimation temperature. The observations of several AGN show a nearly flat NIR SED extending from 2$-$10 $\mu$m \citep{1987ApJ...320..537B}. Some AGN also show silicate features at 10 $\mu m$ \citep{Deo2009} and an excess at $\sim$30 $\mu$m \citep{2011MNRAS.414.1082M}. 
According to AGN unification schemes, this hot dust emission in the NIR can be directly observed in Type 1 sources, whereas in Type 2 sources, it may be obscured by the torus. However, it was reported that some Type 2 AGN  also show NIR excess, which was then attributed to the non-zero probability of radiation emitted by hot graphite dust in the sublimation region of the torus reaching the observer even for the edge-on orientation \citep{0067-0049-204-2-23}. From this, it was inferred that the torus medium could be clumpy. The origin of this observed NIR excess is not understood until today due to lack of information about the sublimation zone. To explain the flat NIR SED in terms of thermal dust emission, a broad range of temperatures from 1000 K to 1800 K is required. 
Isothermal dust close to its evaporation temperature produces a narrower bump \citep{2012MNRAS.423..464H}, but a clumpy torus \citep{2008A&A...482...67S, 2008ApJ...685..147N} allows dust grains of different temperatures to co-exist in the torus, thereby producing a flat NIR SED.
Though it is very difficult to resolve the innermost radius of the torus \citep{refId01}, using advanced techniques like reverberation mapping and interferometry, 
it has been determined for some Seyferts \citep{refId1}. The observed values for the sublimation radii for Seyfert Type 1 AGN are presented in \cite{Gandhi:2015vta}. 
The hottest dust lies in this sublimation region of the torus and could extend up to a radius of 0.1 pc from the nucleus \citep{0004-637X-698-2-1767} and is responsible for the NIR excess in the AGN SED \citep{1978ApJ...226..550R}. This NIR excess for many AGNs was modelled by \cite{1987ApJ...320..537B} using hot graphite dust grains at their sublimation temperature. The presence of hot dust close to its sublimation temperature was later confirmed (for AGN NGC 4151 ) observationally by \cite{minezaki2003inner} using time lag analysis. Dust reverberation measurements are one of the standard techniques used to measure the innermost radii of AGN. Using this technique, \cite{suganuma2006reverberation} and \cite{minezaki2003inner} showed that the NIR emission in the AGN SED is dominated by thermal hot dust components rather than from the accretion disk. 

Various morphologies have been proposed to explain the observed IR SED of AGN, which includes a smooth dust density distribution \citep{1992ApJ...401...99P, 1994MNRAS.268..235G, 1995MNRAS.273..649E, 2003A&A...404....1V, 2005A&A...437..861S} for the torus. A smooth dust distribution could account for the IR SED of AGN but dust grains that are uniformly distributed cannot survive in the near vicinity of an AGN \citep{1988ApJ...329..702K}. The temperature of the dust decreases monotonically with distance from the center. Hence, to prevent dust destruction and for dust grains with different temperatures to coexist at different distances from the centre, models with a clumpy density distribution have been proposed \citep{2007A&A...474..837T, 2008ApJ...685..147N, Thompson_2009, refId01}. Both the smooth and clumpy models of the torus produce similar IR spectra \citep{1987ApJ...320..537B} and can reproduce some of the observations of AGN \citep{1988ApJ...329..702K}. Clumpy torus models \citep{2008ApJ...685..147N} have been known to reproduce the NIR part of the observational SED reasonably well and account for the excess in the NIR, whereas the smooth density models have been able to account for the deep silicate absorption features. Other models like the $CAT3D-wind$ model by \cite{2017ApJ...838L..20H} predict the NIR emission to be from the accretion disk itself and the MIR emission from an outflowing dusty wind. In \cite{Zhuang_2018} the clumpy torus model ($CLUMPY$ from \cite{2008ApJ...685..147N} with the blackbody component) fitted well with the observation for hot dust emission. The clumpy torus model by \cite{Nenkova_2010} was unable to explain the NIR excess in the observed SED of the Seyfert sample of \cite{Liraetal2013}. \cite{selfcon} modelled the NIR emission in AGN using a homogenous disk component with a combined smooth and clumpy density distribution, known as the two phase torus model. In 2012, \cite{doi:10.1111/j.1365-2966.2011.19775.x} proposed the $SKIRT$ model for AGN, consisting of such a two phase torus. This code was used by \cite{stalevski2019dissecting} to model the MIR emission from polar dust in Circinus AGN, which is a Seyfert Type 2 AGN.

The presence of hot dust very close to the central accretion disk has been confirmed by observations of Type 1 AGN. Of the variety of dust grains, silicates and graphites can survive in the AGN torus. Graphite can even survive in the innermost regions of the torus due to its higher sublimation temperature (T $\sim $1800K). Hot dust was thought to be outside the broad line region (BLR) for the AGN, NGC 4151 \citep{1982ApJ...262..564F}. In fact, it was found that pure graphite dust is localised between the dust free BLR and the torus \citep{2009ApJ...705..298M, 2041-8205-737-2-L36, 2012MNRAS.420..526M} and it dominates the SED in the 2-5 $\mu$m region over the power law emission from the disk. It is a challenging task to determine the actual location of the sublimation zone for AGNs due to the compactness of the inner region 
of torus. Also, the existence of the dusty torus, its structure and origin are still debated \citep{Hick2018}. 

Hence, it is crucial to model the NIR AGN SED associated with the innermost regions of AGN using different geometries and compositions of dust and derive the distribution of this hot dust in AGN. A variety of radiative transfer models have been proposed to explain the observed SED mainly focussing on short wavelength emission \citep{gonzalez2019exploring}. 
 In the present work, we apply the $SKIRT$ model to predict the NIR SED of NGC 4151 and thereby derive its central dust distribution and composition. The focus of the present work is to investigate the sublimation zone which may be responsible for the observed NIR excess. We compile the observed IR flux for NGC 4151 from the literature to construct the observed AGN SED which is explained in Section 2. \citet{1999AstL...25..483O} used numerical simulations of the time delay distribution and concluded that the NIR(1-5 $\mu$m) emission region in NGC 4151 cannot be spherically symmetric, but must have the shape of a thin ring or disk. 
Recent observations in the MIR show substantial evidence for the presence of polar dusty winds to be the main source of the MIR emission and a thin disk like region surrounding the central accretion disk to be the source of the NIR emission in AGN rather than a dusty torus \citep{2017ApJ...838L..20H}. This polar structure is thought to surround the narrow line region (NLR). Stalevski et al. (2019b) have successfully modeled the MIR interferometric observations of Circinus AGN using such a polar wind model with a thin disk.  

This is the motivation for our current work. Therefore, apart from the regular two phase torus geometry, we include a thin ring-like layer consisting of graphite grains to account for the NIR SED in NGC 4151. The method and the details of the model along with the model parameters used are explained in Section \ref{sect:Model}.\\

A brief introduction of our target of study, NGC 4151 is presented in the next section. 

\subsection{NGC 4151}
\label{sec:ngc4151}
The target NGC 4151, which is called `The eye of Sauron', is located at a distance of about 19 Mpc \citep{2014Natur.515..528H} and is a well studied AGN. It is a Seyfert Type 1.5 AGN \citep{10.1093/mnras/176.1.61P} having bolometric luminosity $10^{43}-10^{44}$ erg\, s$^{-1}$.
 The SED of Seyfert 1.5 AGN is typically identical to that of Type 1 Seyfert \citep{2009ApJ...702.1127R}. These intermediate Seyfert AGN have a low half-opening angle 
 for the torus ($25^{\circ}$ - $45^{\circ}$) (see Fig. \ref{model}), low values of optical depth and low inclination angle ($i < 50^{\circ}$) \citep{2009ApJ...702.1127R}. \\

It is also a highly variable AGN. It is one of the brightest known AGNs at X-ray wavelengths. The X-ray luminosity is found to be $L_{2-10keV}=1.47\times10^{43}$ erg\, s$^{-1}$ from \textit{Chandra Advanced CCD Imaging Spectrometer (ACIS)} observations \citep{Yang2001} and  $L_{13.6-100keV}=6.2\times10^{43}$ erg\, s$^{-1}$  from Swift-BAT 105-month catalog observations \citep{Oh_2018}.

The supermassive black hole at the center of NGC 4151 has a mass of about 50 $\times$ 10$^{6}$ M$_{\odot}$(\cite{Bentz_2006,2014Natur.515..528H}). It has a weak radio jet having luminosity $\sim 3.87\times10^{37}$ ergs\, s$^{-1}$ \citep{Ulvestad1998}. It contains a radio jet of length $\sim$ 230 pc at position angle $77^{\circ}$ \citep{wilson1982radio}. The presence of a very compact infrared source in the nucleus of NGC 4151 was first suggested by \cite{penston1974broadband}. \cite{swain2003interferometer} concluded that the IR bump in the observed SED of NGC 4151 is due to thermal gas from the accretion disk rather than from the dusty torus. But using NIR spectra obtained with the Gemini Near-Infrared Field Spectrograph(NIFS), \citet{0004-637X-698-2-1767} found that the IR bump is indeed due to the presence of hot dust in the torus and not due to the accretion disk. They derived a temperature of $\sim$  $1285\pm50$K for this hot dust, after subtracting the power-law component due to the accretion disk. \cite{minezaki2003inner} observed the innermost radius of the torus to be $\sim$0.04 pc which is well outside the BLR. This was also in agreement with the size of an unresolved source of NIR emission observed by \cite{Burtscher_2009}. 

\cite{koshida2009variation} reported that there was a strong dependence of the inner radius of the torus of NGC 4151 with the strength of the UV/optical flux from the 
accretion disk. This was attributed to the destruction of grains due to the expansion of the dust sublimation radius when the UV/optical luminosity increases. It was concluded that, when the accretion disk luminosity reduces there must be re-formation of dust grains. 
The self consistent AGN torus model \citep{1988ApJ...329..702K} and interferometric observational studies \citep{pott2010luminosity} give differing views about the location of hot dust in the innermost region of torus. The location of hot dust depends on the AGN luminosity as R$_{\rm sub}$ $\varpropto$ $\sqrt(L_{\rm AGN})$. It should be noted that R$_{\rm sub}$ does not strictly follow the nuclear luminosity always but rather it depends on the high activity state of the nucleus\citep{pott2010luminosity}. From NIR reverberation mapping, the size-luminosity scale is followed which suggests that the emission is dominated by hot dust \citep{suganuma2006reverberation,refId0}. But this empirical scale relation does not hold for MIR emission \citep{refId02}. Hence, the location of hot dust can primarily be determined from the NIR emission. \citet{schunlletal2013} used reverberation mapping for NGC 4151 
as a tool to study the dust distribution and its temperature evolution and concluded that though the NIR emission varied with the accretion disk brightness, the location of hot dust remained static and there was no evidence for dust re-formation.
The outer radius of NGC 4151 torus was estimated to be less than 35 pc \citep{Radomski2003} which is consistent with other measurements of the size of the torus in the literature \citep{2003MNRAS.340..733R, 1990AJ.....99.1456N}.

\section{OBSERVED SED}
\label{sect:data}
To compare the model for NGC 4151 with observations,
we have taken the observational data from \cite{2003AJ....126...81A}. They had derived the SEDs in the range
0.4 $\mu$m-16 $\mu$m for an extended sample of Seyferts consisting of 58 galaxies including NGC 4151.
They have used high spatial resolution data in the wavelength range from 0.4 $\mu$m to 21 $\mu $m corresponding to I, J, H, K, L, M, N and Q bands. 
Specifically observations in the range, 1.1-2.2 $\mu$m, are from $HST/NICMOS$ and for 3.8-4.8 $\mu$m range, the data are taken from ground-based UKIRT(United Kingdom Infrared Telescope) observations. The MIR data points are with a small aperture.  The unresolved flux in the optical, NIR and MIR range has an uncertainty of $\pm$30\%. Most of the errors arise due to the background subtraction.
In addition, the data for 5.5$\mu$m, 7.7$\mu$m and 16$\mu$m are from $Infrared~Space~Observatory$(ISO). 
The $ISO$ data has higher sensitivity and the nuclear-unresolved luminosity has not been subtracted. 
In the observed data, the nuclear spectrum itself contains little or no PAH emission. Seyfert 2 galaxies show PAH features that could be associated with the host galaxy or star formation as PAHs are good tracers of star formation in galaxies \citep{Peeters_2004}. PAH contributions have not been subtracted from the SED of NGC 4151 as the star formation rate is weak in the circumnuclear region \citep{2008ApJ...681..141R}. For NGC 4151, the AGN emission dominates the stellar emission in the nuclear region. These data are compared with our proposed model SED to infer the geometry of the dusty torus, where silicate and graphite dust are the main sources of IR emission since PAH sublimate very easily at the temperatures that exist close to the central nucleus.

\begin{table}
\begin{center}
\caption[]{SED of NGC 4151 for optical and infra-red fluxes}
\label{obsflux}

 \begin{tabular}{clcl}
 \hline\noalign{\smallskip}
No &  Wavelength($\mu $m)       & Flux(W~ m$^{-2}$)      \\
 \hline\noalign{\smallskip}
1  & $ 0.8 $(I band)      & $1.92\times10^{-13}$        \\
2  & $ 1.2 $(J band)      & $1.73\times10^{-13}$         \\
3  & $ 1.6 $(H band)      & $1.87\times10^{-13}$              \\
4  & $ 2.2 $(K band)      & $2.45\times10^{-13}$              \\
5  & $ 3.5 $(L band)      & $2.82\times10^{-13}$            \\
6  & $ 4.8 $(M band)      & $2.84\times10^{-13}$          \\
7  & $ 5.5 $              & $4.81\times10^{-13}$         \\
8  & $ 7.7 $              & $4.24\times10^{-13}$         \\
9  & $ 10.6$(N band)      & $3.77\times10^{-13}$       \\
10 & $ 16  $              & $7.72\times10^{-13}$       \\
11 & $ 21  $(Q band)      & $4.57\times10^{-13}$       \\

  \noalign{\smallskip}\hline
\end{tabular}
\end{center}
\end{table}

\begin{figure}
  \centering
   \includegraphics[width=90mm]{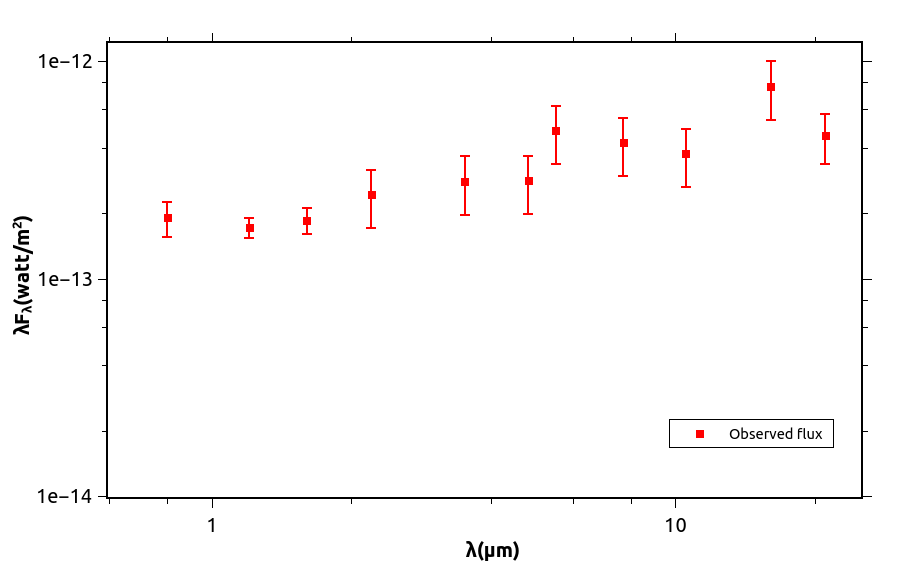}
	  \caption{\small The observed SED of NGC 4151}
\label{observed}
\end{figure}
The observed flux of NGC 4151 are given in Table \ref{obsflux} and plotted in Fig. \ref{observed}. 

\section{METHOD AND MODEL}
\label{sect:Model}
 The SED of NGC 4151 shows NIR excess emission which could arise due to the presence of hot dust close to the central nucleus. We have attempted to model this excess emission by incorporating a region of hot dust grains close to their sublimation temperatures in the innermost 
 regions around the accretion disk of NGC 4151.
The sublimation temperature of silicates is around 1200 K and that for the graphite is around 1800 K. Hence for a mixture of silicate and graphite grains that follow the MRN size distribution \citep{1977ApJ...217..425M}, the average sublimation temperature would be $\sim$ 1500 K. Therefore, apart from a model with a single dust sublimation radius (which is also the innermost radius of the torus), we also consider a model with a sublimation zone; i.e. an inner ring-like region with only graphite dust at a temperature of 1800 K and an outer torus that contains a mixture of silicate and graphite dust grains. To model the observed SED of NGC 4151, we use the $SKIRT$ code \citep{steinacker2013three, doi:10.1046/j.1365-8711.2003.06770.x, 0067-0049-196-2-22} which solves the radiative transfer equation for dusty astrophysical systems. It is a three-dimensional radiative transfer code, which employs the Monte Carlo method. The inputs to the code are, the incident luminosity in the form of photon packets, the inner (R$_{\rm in}$) and the outer (R$_{\rm out}$) radius of the torus, its optical depth ($\tau_{9.7}$), the half opening angle ($\sigma $), the number of 
clumps ($N_{\rm clump}$) and the clump size ($R_{\rm clump}$) in the torus. The torus IR SED 
consisting of the variation of flux with wavelength is the output of the code. The advantage of the $SKIRT$ code is that no iterative procedure is necessary for the model as it simultaneously accounts for the primary emission, dust absorption and dust emission from each dust cell.

\subsection{Description}

The model can accommodate different types of geometries like a torus geometry, a ring geometry or a combination of both together with a central source. The accretion disk acts as the primary source of radiation, surrounded by the dusty torus, where the dust absorbs most of the UV/optical photons of the incident radiation from the disk and re-emits in the IR. The geometry of the torus is defined by its inner radius, outer radius and the half opening angle. The morphology of the torus is characterised by specifying its optical depth, the number of clumps and the dust distributions. The inner radius of the sublimation zone is determined by the sublimation temperature of the dust grains irradiated by the central accretion disk near the boundary at which, the dust starts to sublimate. Hence, the inner radius of the torus is
considered to be the sublimation radius. For the silicate-graphite mixture, the grain size in the torus ranges from very small 0.005 $\mu$m sized grains to larger 0.25 $\mu$m sized grains. The sublimation radius of the torus for an average grain size of 0.05 $\mu$m for the mixture is given by \citep{1987ApJ...320..537B} as\\
\begin{equation}        
	R_{\rm sub}^{\rm Si,\rm Gr} = 1.3 \sqrt{L^{\rm AGN}_{46}}~ T^{-2.8}_{1500}~ a^{-0.5}_{0.05}~~ \rm {pc}
\end{equation}
The sublimation radius of the graphite alone for the average grain size of 0.5 $\mu$m \citep{doi:10.1093/mnras/stx2850} is given by 
\begin{equation}
	R_{\rm sub}^{\rm Gr}=0.5 \sqrt{L^{\rm \scriptsize AGN}_{46}}~T^{-2.8}_{1800}~ a^{-0.5}_{0.5}~~ \rm{ pc}
\end{equation}
where $ L^{\rm AGN}_{46} $ is the AGN luminosity in units of $10^{46}$\,erg\, s$^{-1}$. The luminosity, $L^{\rm AGN}_{46}$ is the central bolometric luminosity which is absorbed by the dust grains and re-radiated in the IR through the radiative transfer mechanism. The parameter $T_{1500}$ is the average grain sublimation temperature for a mixture of silicate and graphite grains and $T_{1800}$ is the sublimation temperature of pure graphite dust grains.

\begin{figure}
\centering
\includegraphics[width=\columnwidth]{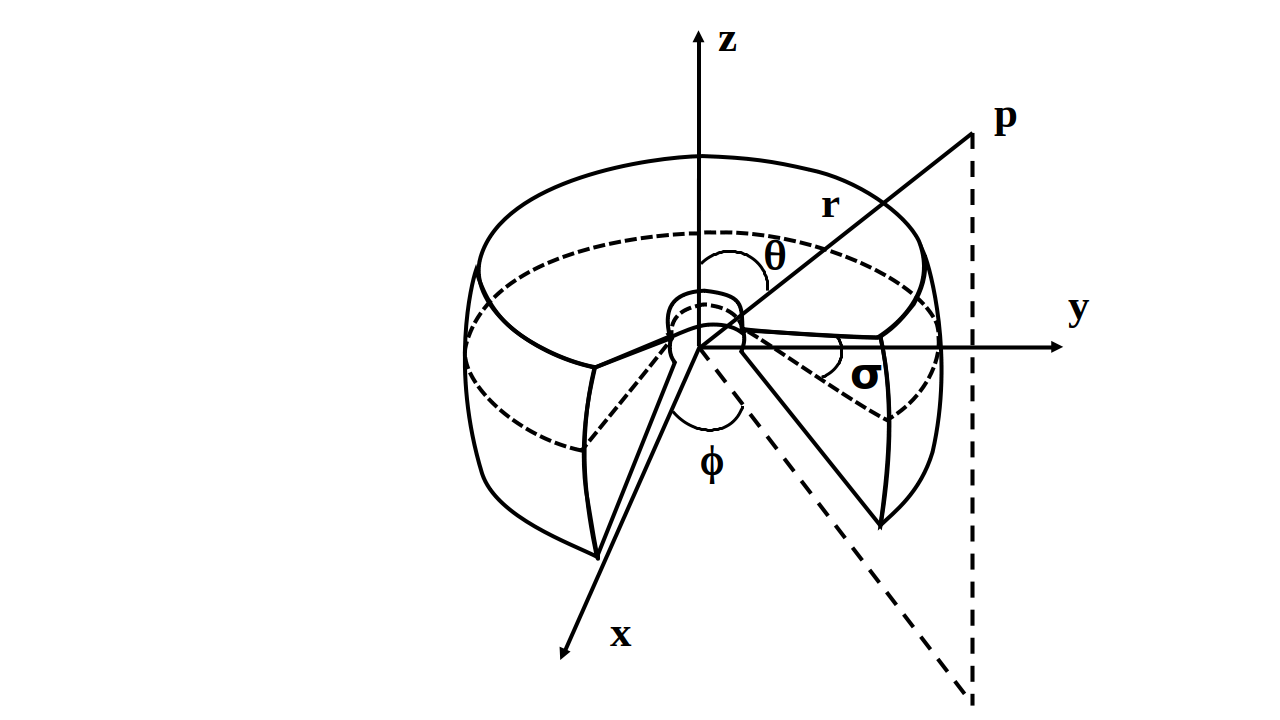}
\caption{Schematic diagram of the adopted geometry where $\theta$ is the polar angle, $\sigma$ is the half opening angle 
	and $\phi$ is the azimuthal angle \citep{doi:10.1111/j.1365-2966.2011.19775.x}.
}
\label{model}
\end{figure}

The inner radius of the torus $R_{\rm in}$, in general can depend on the polar angle $\theta$ (see Fig. \ref{model}) due to the disk luminosity being anisotropic \citep{ doi:10.1093/mnras/stw444}:
\begin{equation}
R_{\rm in}= R_{\rm iso}[\rm cos\theta(2\rm cos\theta+1)]
\end{equation}

The dust distribution that allows a density gradient along the radial direction $r$ and the polar angle, $\theta$ is given
by 
\begin{equation}
 \rho(r,\theta) \varpropto r^{-p}e^{-qcos\theta}
	\label{dustdist}
\end{equation} 
where r and $\theta$ are the polar coordinates. Here, $p$ is the polar index and $q$ is the polar exponent \citep{ doi:10.1093/mnras/stw444}.\\

The grain size distribution in the present model is given by 
\begin{equation}         
 dN(a)=10^{A_i}a^{q'} da    \\
\end{equation}
where $dN(a)$ is number of grains with size between $a$ and $a+da$ normalized to the number of hydrogen atoms \citep{1977ApJ...217..425M}. The value of $-3.5$ for exponent $q'$ is taken from \cite{1984ApJ...285...89D}. The constant $A_i$ is normalized to the hydrogen abundance. It is $-25.16$ for graphite and $-25.11$ for silicate \citep{1984ApJ...285...89D}. 

Dust is distributed in a 3D Cartesian grid composed of a large number of cubic cells, where we consider 100 cells along each axis for the current simulation. The flux is calculated for wavelengths ranging from 0.1$\mu$m to 180  $\mu$m.  
The parameters for NGC 4151 that are fixed in the present model are described below and summarized in Table \ref{para_model}.

\begin{table*}
\begin{center}
\caption[]{The fixed parameters for all the models}
\label{para_model}
\begin{tabular}{cccc}
\hline 
Adopted parameters &  value & Reference \\ 
\hline 
$L_{Bol}$ & 10$^{10}$L$\odot$ &  \citep{2002ApJ...579..530W} \\  
$R_{\rm out}$ & 15 pc & \citep{ doi:10.1093/mnras/stw444} \\  
$A_{i}$ and $q'$    &  -25.16(for graphite), -25.11(for silicate) and -3.5 & \citep{1984ApJ...285...89D} \\
 $p$(polar index)               &     1    &   \citep{ doi:10.1093/mnras/stw444}      \\
 $q$(polar exponent)               &      0     &   \citep{ doi:10.1093/mnras/stw444}   \\
Filling factor    &      0.25     &  \citep{ doi:10.1093/mnras/stw444}  \\
$F_{\rm clumps}$        &       0.97  &    \citep{ doi:10.1093/mnras/stw444}      \\
$N_{\rm clumps}$       &           $\sim$ 8000 &   \citep{ doi:10.1093/mnras/stw444}     \\
Contrast                &     100    &  \citep{ doi:10.1093/mnras/stw444} \\
\hline
\end{tabular}
\end{center}
\end{table*} 

The parameter $L_{\rm Bol}$ represents the bolometric luminosity of the AGN.  The outer radius, $R_{\rm out}$ of the torus is assumed to be 15 pc. 
 The total amount of dust is determined by the equatorial optical depth at 9.7$\mu$m which is used as an input to the algorithm that generates the clumps inside the geometry. After applying the algorithm, dust is distributed into the clumpy two phase medium. The two phase medium consists of a large number of high density clumps embedded in a smooth low density medium. So, the optical depth along the line of sight varies according to the number of clumps \citep{doi:10.1111/j.1365-2966.2011.19775.x} thereby changing the SED. 
 For the present model, the dust distribution parameters in Eqn. \ref{dustdist} have been assigned the values, $p=1$ and $q = 0$ \citep{doi:10.1093/mnras/stw444}. 
 
 The parameter $N_{\rm clump} $ is the number of clumps inside the torus where grains are present in a high density medium and $F_{\rm clump} $ is the fraction of total dust mass in the torus that has been
locked up in the clumps. The remaining dust is distributed in the torus in the form of a smooth medium. 
The number and size of the clumps are usually chosen to achieve a certain volume filling factor of the torus. Apriori, the number of clumps and the outer radius of the torus are not known. 
As we have adopted the Stalevski torus model, the number of clumps has been fixed at $\sim$ 8000, irrespective of the size of the torus. As a result, they could overlap with each other and form high density complex structures. The contrast parameter defined as the ratio of the high and low densities of the medium, is kept constant at 100.
\\

\subsection{Torus only model and RAT model} 
We have considered two models to explain the IR SED of NGC 4151, i.e. the `Torus Only' model (TO model hereafter) which consists only of a torus geometry and the `Ring And Torus' model (RAT model hereafter) which consists of a ring of graphite dust surrounded by the torus. In the RAT model, we perform two tests. In one test, the ring is considered to have a smooth dust distribution (`smooth RAT' model) and in the other, it is considered to be clumpy (`clumpy RAT' model). The models are explained in detail in the next section.
\subsubsection{Torus only model}
In the TO model the torus consists of a two phase smooth and clumpy medium. The accretion disk emits a bolometric luminosity $\sim$ $10^{43}$ erg\, s$^{-1}$ \citep{2002ApJ...579..530W, Penston611963} which is partially absorbed by the dust in the torus. 
We have studied the variation of the TO model SED with the half-opening angle ($\sigma$) of the torus, the inner radius ($R_{\rm in, \rm t}$), the inclination angle 
($i$), the distance ($d$) from the observer, the optical depth ($\tau_{9.7, \rm t}$) of the dust in the torus, and the
clumpsize (R$_{\rm clump}$). We have considered three different distances for NGC 4151, 13.3 Mpc \citep{10.1046/j.1365-8711.1999.02331.x}, 19 Mpc\citep{2014Natur.515..528H}and 29 Mpc\citep{Yoshii_2014}. The inclination angle $i$ is also a parameter in the present model as the observed SED largely depends on the angle at which
we view the source.

The MRN size distribution is used for the grains in the torus of this model with grain sizes ranging from 0.005 $\mu$m to 0.25 $\mu$m for the silicate-graphite mixture. 
The dust sublimation radius ranges from 0.02 pc to 0.13 pc for the above grain size range, of average sublimation temperature 1500K.
 The inner radius takes the values 0.04 pc, 0.06 pc and 0.1 pc and the values of optical depths are taken as 0.1,1,5,10 and 15. 
 The details of the parameters of the torus that are varied in the TO model are presented in Table \ref{pr}.
 
\begin{table}
\begin{center}
\caption{Parameters varied in TO model}
\label{pr}
\begin{tabular}{clcl}
  \hline\noalign{\smallskip}
S.No. &  Parameters       & Adopted values      \\
  \hline\noalign{\smallskip}

Torus                &                 \\
\hline

1  & Inclination angle(i) & 0$^\circ$-90$^\circ$ \\
2  & $R_{\rm in, \rm t}$         &   0.04 pc,0.06 pc,0.1 pc  \\
3  & $\sigma$      & 20$^\circ$, 25$^\circ$,30$^\circ$ and 50$^\circ$  \\
4  & Grain size         & 0.005 $\mu$m-0.25 $\mu$m  \\ 
5  & $R_{\rm clump}$    & 0.04 pc, 0.4 pc and 0.2 pc \\
6  & $\tau_{9.7, \rm t}$             & 0.1, 1, 5, 10, 15 \\
7  & Rest of the parameters are &  \\
   & same as in table 2.        &  \\   
  \noalign{\smallskip}\hline

\end{tabular}
\end{center}
\end{table}

Since pure graphite dust is considered to be a possible source of the NIR excess in AGN, a separate NIR emitting region can be incorporated in the model, as a
sublimation zone consisting of pure graphite grains located between the accretion disk and torus. However, in the TO model, it is not feasible to introduce such a sublimation zone. Hence we progress to the RAT model where the sublimation zone is introduced as an inner graphite ring surrounding the accretion disk \citep{doi:10.1093/mnras/stw444, 2012MNRAS.420..526M}.

The details of the RAT model are explained in the next section. 

\subsubsection{Clumpy RAT model} 
The motivation for introducing pure graphite component into the model is primarily due to the results of \cite{2009ApJ...705..298M, 2041-8205-737-2-L36, 2012MNRAS.420..526M} where it was shown that such a component is necessary in order to explain the NIR emission of Type I QSOs. Dust grains when irradiated by UV/optical photons get heated and emit thermal radiation at a longer wavelength. When the grains are sufficiently large in size, this radiation is expected to be in the NIR range. For large silicate grains, the grain temperature is higher than the sublimation temperature. So, large silicate grains cannot exist at very high temperatures $\sim$ 1800 K. However, pure graphite dust can survive at such temperatures. The location of pure graphite dust must be external to the dust free BLR, and internal to the region where the standard silicate-graphite mix type grains exist.

\begin{figure}
\centering
\includegraphics[width=\columnwidth]{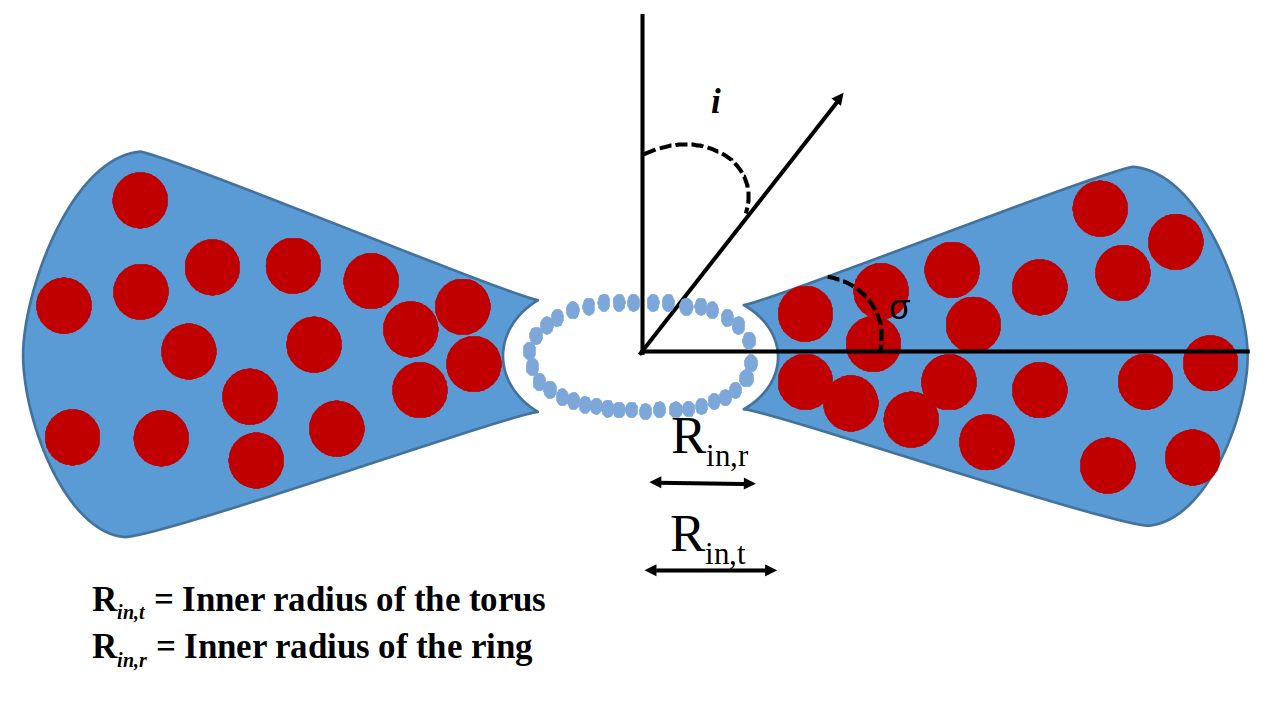}
\caption{Schematic diagram of the cross sectional view of the RAT model with clumpy ring: `$\sigma$' is the half opening angle and `$i$' is the inclination angle.}
\label{sdiag}
\end{figure}

In this model, the sublimation zone consists of a dusty torus with an MRN dust size distribution, and the ring consisting of only graphite particles in the region with sizes ranging from 0.1 $\mu$m to 1 $\mu$m as only large grains can survive in the innermost region of the torus \citep{doi:10.1093/mnras/stx2850}. Hence, the sublimation radius of pure graphite ring ranging from 0.01 pc to 0.06 pc correspond to the large grain sizes and of temperature 1800 K. In the current model, the torus geometry is detached from the ring geometry and hence inner radius of the torus does not coincide with the outer radius of the ring. The geometry of the torus part is retained as in the TO model. The schematic diagram of the cross-sectional view of the RAT model with the clumpy ring is shown in Fig. \ref{sdiag}. The details of the parameters for clumpy RAT model are presented in Table \ref{tr}. The following are the properties of dust in the clumpy RAT model. \\

\begin{itemize}
	\item The inner radius and the optical depth of the ring, $R_{\rm in,r}$ and $\tau_{9.7, \rm r}$ are varied as free parameters while the inner radius of the torus and its optical depth, $R_{\rm in,t}$ and $\tau_{9.7,\rm t}$ are fixed. The width and height of the ring are kept constant and are considered to be very thin in comparison to that of the torus.
\item The pure graphite composition, incorporated in the ring, follows the MRN dust size distribution \citep{1977ApJ...217..425M}. 
\item The dust distribution in the ring is considered to be present in a clumpy form. The parameters for the torus geometry, dust density and the dust composition remain the same as in the TO model.
\item Two different clump sizes have been considered i.e. 0.002 pc and 0.04 pc in the present model. 
	
\end{itemize}

\begin{table}
\begin{center}
\caption[]{Parameters in clumpy RAT model}
\label{tr}
\begin{tabular}{clc}
  \hline\noalign{\smallskip}
S.No. &  Parameters       & Adopted values      \\
\hline
	&   Ring                     &               \\
\hline
1 & $R_{\rm in,r}$         &    0.01pc -0.06 pc       \\
2 & Width          &   0.02 pc             \\
3 & Height        &   0.04 pc       \\
4 & $\tau_{9.7, \rm r}$  &  0.1-100 \\
5 & $f_{\rm clumps}$     &   0.97            \\
6 & $N_{\rm clumps}$     &   $\sim$ 4323 \\
7 &  $R_{\rm clump,r}$ & 0.002 pc, 0.04 pc \\
8 & $R_{\rm out,r}$        &   $R_{\rm in, r}$ + Width \\
9 & Grain size         & 0.1 $\mu$m-1 $\mu$m \\
\hline\noalign{\smallskip}
	&    Torus               &              \\
\hline
1  & $R_{\rm in,t}$         &   0.1 pc \\
2  & $\tau_{9.7, \rm t}$     &   10                           \\
3  & Grain size         & 0.005 $\mu$m-0.25 $\mu$m  \\
	4  & $R_{\rm clump}$ & 0.4 pc  \\
5  & $\sigma$           & 30$^{\circ}$, 50$^{\circ}$ \\
6  & Rest of the parameters are                                    \\
   & same as in Table 2.  & \\
\hline
\end{tabular}
\end{center}
\end{table}

\subsubsection{Smooth RAT model}
\begin{figure}
\centering
\includegraphics[width=\columnwidth]{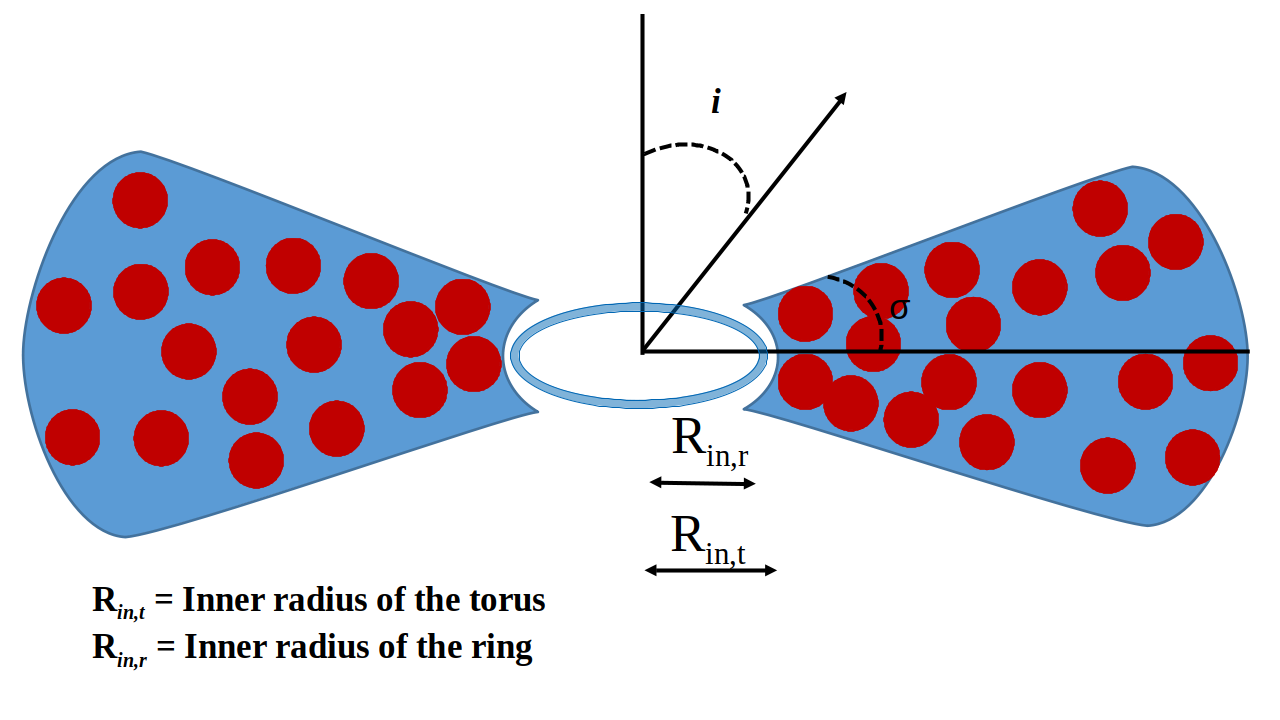}
\caption{Schematic diagram of cross section of RAT model with smooth ring: `$\sigma$' is the half opening angle and `$i$' is the inclination angle.}
\label{smoothrat}
\end{figure}

The smooth RAT model consists of an accretion disk, a ring with smooth distribution of dust and a clumpy torus. The torus and ring geometry are separated from each other like clumpy RAT model. The schematic diagram of the cross-sectional view of the RAT model with the smooth ring is shown in Fig. \ref{smoothrat}. 
The details of the parameters for smooth RAT model are presented in Table. \ref{rr}. \\

\begin{itemize}
\item The inner radius, height and width of the ring are same as for clumpy RAT model.
\item The torus geometry in this model is fixed as in TO and the clumpy RAT model.
\item The inner radius and optical depth of the ring are the free parameters in this model too.
\end{itemize}

\begin{table}
\begin{center}
\caption[]{Parameters in smooth RAT model}
\label{rr}

\begin{tabular}{clcl}
  \hline\noalign{\smallskip}
S.No. &  Parameters       & Adopted values      \\
  \hline\noalign{\smallskip}                
	&   Ring                    &                      \\
\hline
1 & $R_{\rm in,r}$       &   0.01 pc-0.06 pc       \\
2 & Width          &   0.02 pc             \\
3 & Height         &   0.04 pc       \\
4 & $\tau_{9.7, \rm r}$   &   0.1-100 \\
5 & $R_{\rm out,r}$        &   $R_{\rm in, r}$ + Width \\
6 & Grain size         &  0.1 $\mu$m-1 $\mu$m \\
 \hline\noalign{\smallskip}
	&    Torus                   &          \\
\hline
1  & $R_{\rm in,t}$         &   0.1 pc     \\
2  & $\tau_{9.7, \rm t}$     &   10                           \\
3  & Grain size         & 0.005 $\mu$m-0.25 $\mu$m  \\
4  & $\sigma$           & 30$^{\circ}$,50$^{\circ}$ \\
5  & Rest of the parameters are                                    \\
   & same as in Table 2.            & \\                     
\hline
\end{tabular}
\end{center}
\end{table}

\section{RESULTS AND DISCUSSION}
In this paper, we have considered three different scenarios for the dust distribution around the accretion disk in order to explain 
the observed SED of NGC 4151 (1) TO model, (2) clumpy RAT model and (3) smooth RAT model. In RAT model, instead of one sublimation radius, two sublimation radii are taken into account. The radiative transfer depends on the geometry and properties of this sublimation zone.

The parameters used in the above models are given in Tables \ref{para_model}, \ref{pr}, \ref{tr} and \ref{rr}. The luminosity of the accretion disk is $\sim 10^{10}$ $L_\odot$ \citep{2002ApJ...579..530W} and the distance of 19Mpc \citep{2014Natur.515..528H} is
obtained by scaling the model SED for different distances with the observed SED (see Fig. \ref{Distance}). 

We have compared our model SEDs with the observed data upto 21 $\mu$m with primary focus on the NIR part of the SED. \\
The parameter estimations are done from the goodness of fit test by two methods i.e. $\chi^2$ test and $R^2$ statistics.

\smallskip
\subsection{MODEL versus OBSERVATIONS}

We present here the results of the `TO', the `clumpy RAT' and the `smooth RAT' models by varying the parameters presented in 
Tables. \ref{pr}, \ref{tr} and \ref{rr} and their comparisons with observed SED.
The goodness of fit for each model is calculated by using the reduced $\chi^{2}$ function, which is defined by \\
$\chi^{2} = \frac{1}{(n-p)}\sum\limits_{i=1}^{n} (\frac{\rm O_{i}- \rm M_{i}}{ \rm \delta_{i}} )^{2}$ \\
where, $\rm O_{i}$ and $\rm M_{i}$ are the observed data and model data for ith photometry point and $\rm \delta_{i}$ is the observational error, $n$ is the number of observations and $p$ is the number of free parameters.
\subsubsection{Torus only model}

\begin{figure}
\centering
\includegraphics[width=\columnwidth]{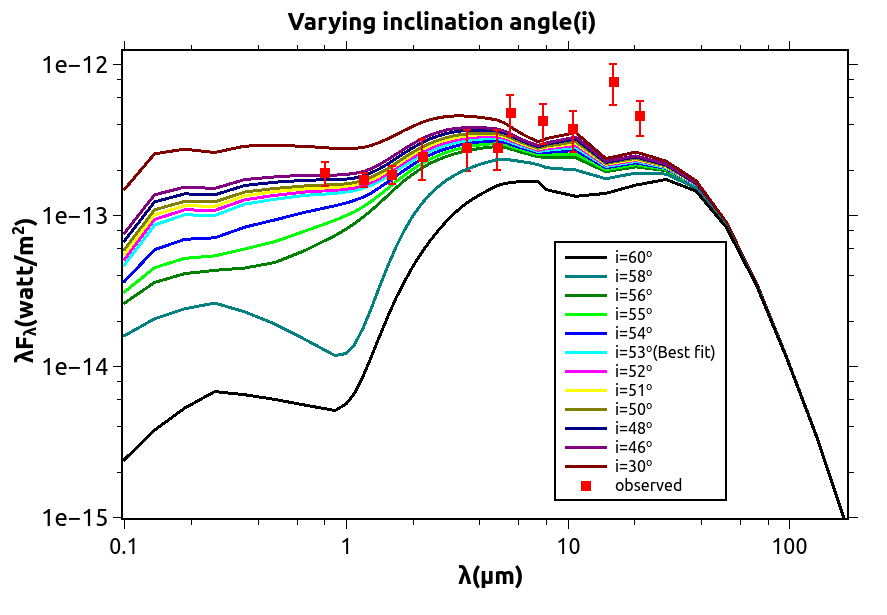}
\caption{The SED with TO model for different inclination angle at half-opening angle 30$^{0}$, $\tau_{9.7, \rm t}$=10, $R_{\rm in, \rm t}$ = 0.06 pc, $R_{\rm out}$ = 15 pc, clump size = 0.4 pc and 
$p$ = 1 and $q$ = 0. 
}
\label{inclination}
\end{figure}

\begin{figure}
\centering
\includegraphics[width=\columnwidth]{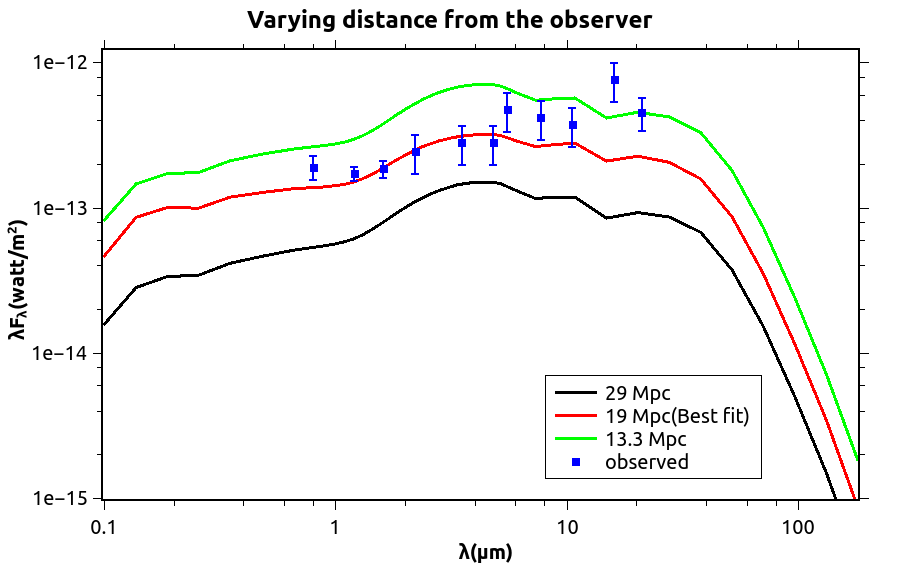}
\caption{The SED with TO model for different distances from the observer for half-opening angle 30$^{\circ}$, $\tau_{9.7, \rm t}$=10, $R_{\rm in, \rm t}$ = 0.06 pc, $R_{\rm out}$ = 15 pc, $i$= 53$^{0}$, clump size = 0.4 pc and $p$ = 1 and $q$ = 0.  
}
\label{Distance}
\end{figure}

\begin{figure}
\centering
\includegraphics[width=\columnwidth]{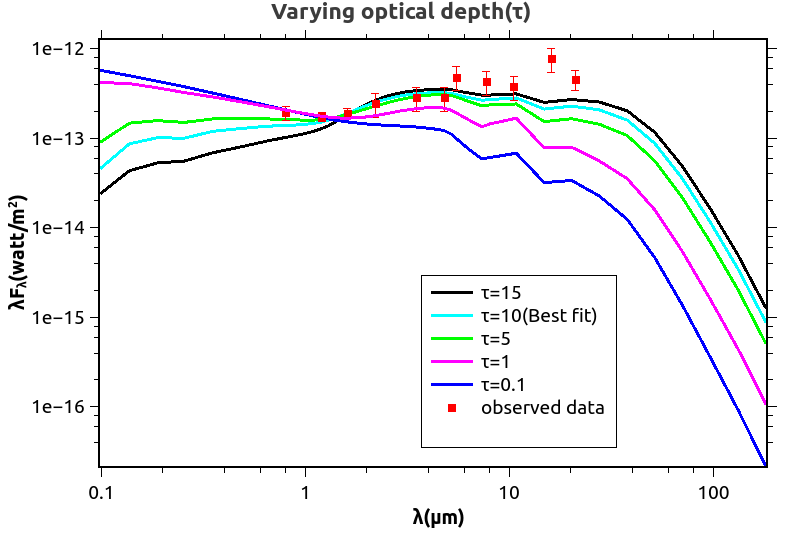}
\caption{The SED with the TO model for different optical depths of the torus with : half-opening angle 30$^{\circ}$, $R_{\rm in, \rm t}$ = 0.06 pc, $R_{\rm out}$ = 15 pc, $i$= 53$^{\circ}$, clumpsize = 0.4 pc and $p$ = 1 and $q$ = 0.
}
\label{optical}
\end{figure}

\begin{figure}
\centering
\includegraphics[width=\columnwidth]{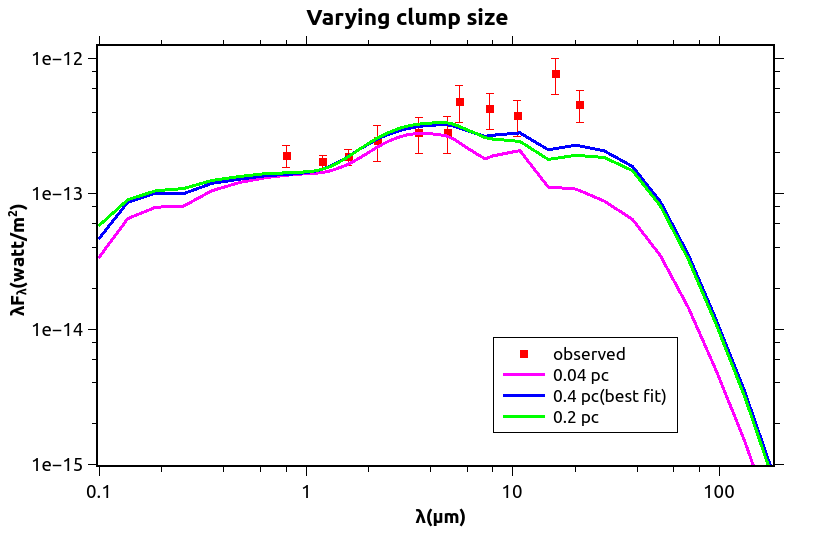}
\caption{The SED with the TO model for different clump sizes with : half-opening angle 30$^{\circ}$, $\tau_{9.7, \rm t}$=10, $R_{\rm in, \rm t}$ = 0.06 pc, $R_{\rm out}$ = 15 pc, $i$= 53$^{0}$ and 
$p$=1 and $q$ = 0.
}
\label{clumpsize}
\end{figure}

\begin{figure}
\centering
\includegraphics[width=\columnwidth]{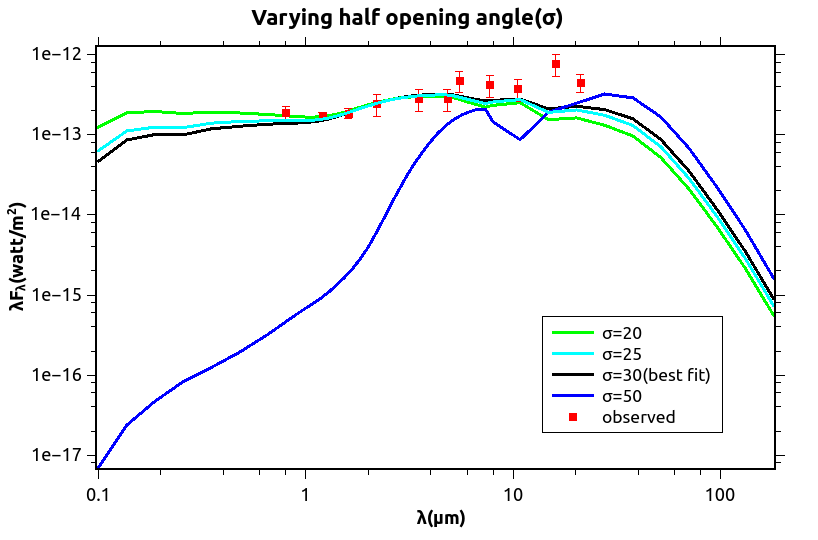}
\caption{The SED with the TO model for different half-opening angles with : $\tau_{9.7, \rm t}$=10, $R_{\rm in, \rm t}$ = 0.06 pc, $R_{\rm out}$ = 15 pc, $i$ = 53$^{\circ}$, clumpsize = 0.4 pc and 
$p$=1 and $q$ = 0.
}
\label{halfopeningangle}
\end{figure}

We focus on the flattened NIR part of the SED profile (see Fig. \ref{observed}) in order to derive the structure and composition of the innermost regions of the torus.  In the AGN unification scheme, the inclination angle is the most crucial parameter.  For NGC 4151, using the reduced $\chi^2$ test, we find from our simulation that the observer's line of sight makes an angle of 53$^{\circ}$ $^{+3}_{-11}$ with the polar axis of the AGN torus as shown in Fig. \ref{inclination}. 
The SEDs obtained for different model parameters are plotted in Figs. \ref{Distance}, \ref{optical}, \ref{clumpsize}  and \ref{halfopeningangle}. 
The Fig. \ref{Distance} shows the variation of SED with the distance to the source. The best fit distance is found to be 19 Mpc.
The best fit values obtained for the other parameters are :
 $\tau_{9.7, \rm t}$=10, $R_{\rm in, \rm t}$ = 0.06 pc, $R_{\rm clump}$ = 0.4 pc. The
flat nature of the NIR SED together with the deduced orientation of 53$^{\circ}$ $^{+3}_{-11}$ (see Fig. \ref{inclination}),  suggests the presence of fewer clouds along this particular line of sight thereby allowing direct observation of the central region of NGC 4151 \citep{2009ApJ...702.1127R}. 

We can see from Fig. \ref{optical} how the optical depth of the torus affects the model SED. The best fit value was
found to be $\tau_{9.7, \rm t}=10$. We note that any further increase in optical depth results in lower emission in UV/optical range because of greater extinction and higher emission in NIR and MIR.
The TO model is investigated for various inner radii 0.04 pc, 0.06 pc and 0.1 pc. The model SEDs for varying clumpsizes and half-opening angles are shown in Figs. \ref{clumpsize}, \ref{halfopeningangle} 
respectively.
The lowest value of $\chi^2$ corresponds to $R_{\rm in, t} = 0.06$pc (see Table. \ref{A2} in Appendix A).


Theoretically, the range of dust sublimation radii vary from 0.02 pc to 0.13 pc corresponding to the silicate-graphite dust mixture with an average sublimation temperature 1500 K.


	 However, if we consider an average grain size of 0.05$\mu $m for the mixture, the sublimation radius is 0.08 pc 
	 which is greater than the best fit radius derived here. 
 Therefore, we fix the best fit parameters for the TO model to be : $R_{\rm in,\rm t}$=0.1 pc with $\sigma$ = 30$^{\circ}$, $\tau_{9.7, \rm t}$ = 10 and $R_{\rm clump}$=0.4 pc. 
 The best fit model SED is plotted alongside the observed SED in Fig. \ref{comp_all}.
 The ratio of outer radius to best fit inner radius of the torus is 150, which is 170 in case of \cite{selfcon}.
 However, the observed inner radius of 0.04 pc determined using Interferometric technique and reverberation mapping, 
 does not match with our best fit radius.

Therefore, we progress to the RAT model to investigate if it can provide a better agreement with the 
observed inner radius and the SED.
 Here we use the best fit parameters from the TO model and incorporate these values for the torus in RAT model.
 The results obtained from RAT model are given in the next section.

\subsubsection{Clumpy RAT model}

Now, we examine the results of the clumpy RAT model, where  clumps of large graphite dust of sizes 0.1-1$\mu$m are present in the ring. 

\begin{figure}
 \centering
   \includegraphics[width=\columnwidth]{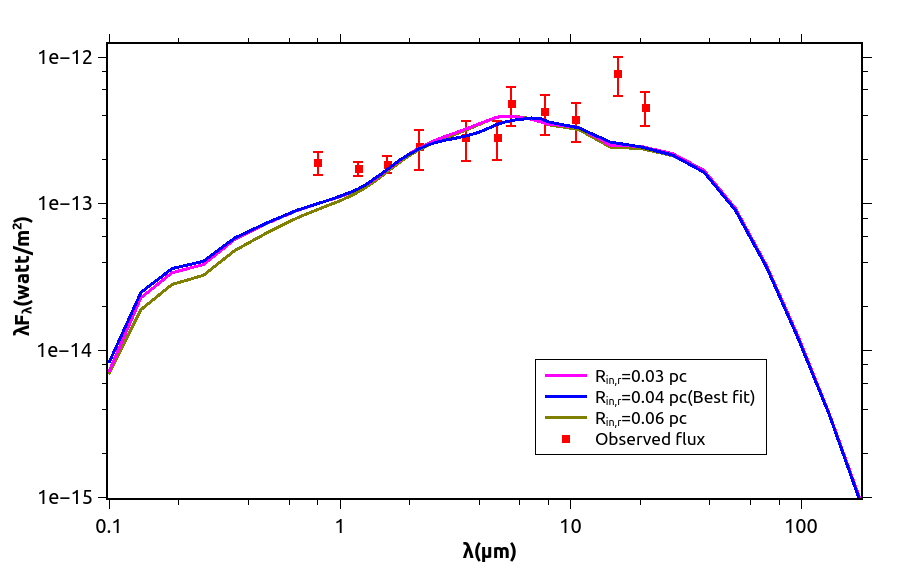}
   \caption{{The SEDs of clumpy RAT model for different ring inner radii.} }
\label{radius_rat}
\end{figure}
\begin{figure}
  \centering
   \includegraphics[width=\columnwidth]{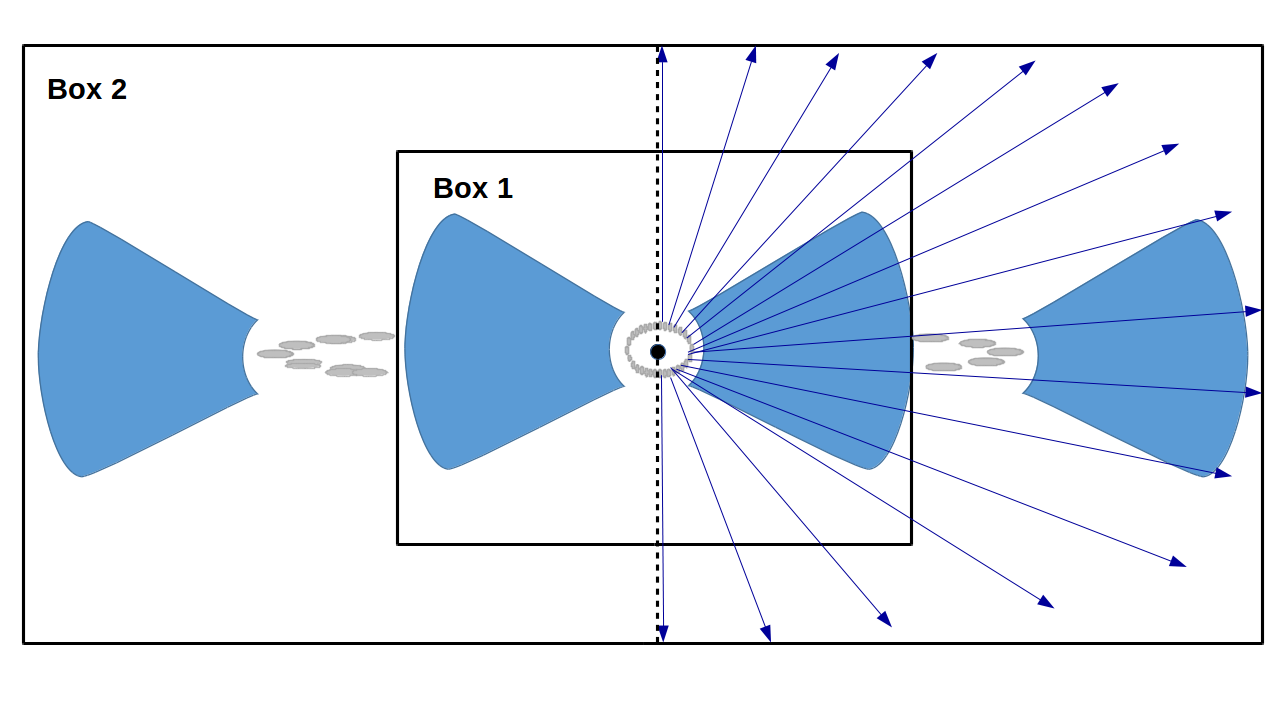}
  \caption{{Schematic diagram of clumpy RAT model with different radii with the source of UV radiation at the centre. Box 1 represents the torus with ring lying at 0.01 pc and  box 2 represents the torus with ring lying at 2 pc. If the radiation is incident on the box 1, the torus is being exposed to more radiation than it would in box 2.}}
  \label{rad_diag}
\end{figure}

 The inner radius of the ring is varied from 0.01 pc to 0.06 pc by keeping torus inner radius at 0.1 pc and optical depth of the ring 
 varied from 0.1 to 100. We have tested the model for $R_{\rm in, t}$ = 0.2 pc to 5 pc and have found that the $\chi^2 $ value increases with increase in
 $R_{\rm in,t}$.
This is because as the torus moves away from the accretion disk, the NIR contribution from the torus reduces significantly. The graphite ring alone 
cannot result in a flat NIR emission. The output SED of this model with different ring inner radii are plotted in Fig. \ref{radius_rat}.
We see that all the models reasonably fit the observed SED, however, the models deviate from the data at optical/NIR wavelengths.
As the inner radius is increased, the amount of radiation incident on the ring decreases. 
This is depicted in Fig. \ref{rad_diag} (not drawn to scale) where more dust is being exposed to the incident radiation in box 1 compared to box 2. 

We have analyzed the model flux of NGC 4151 at an inclination angle ($i$) of 53$^{\circ}$ $^{+3}_{-11}$, derived from the TO model. 
In order to investigate the possibility of a different best fit inclination angle, we treated $i$ as a variable parameter in this model and obtain
the best fit to be 48$^{\circ}$ (see Fig. \ref{fig:figs1} and the $\chi^2$ Table.\ref{tab:tabs1} in Appendix B) with least $\chi^2$ value 1.033. 
In the clumpy RAT model, the visibility of the central source depends on the inclination angle 
and chance of encountering clumps along the 
observer's line of sight (i.e. $\tau_{\rm total}$). The clumpy RAT model explains the observed flux reasonably well
for different inclination angles and they all lie well within the acceptable range derived from the TO model. Hence, 
we proceed with $i =53^{\circ}$ in this model.

The clumpy RAT model SED shows a reduction in flux in the optical/NIR region, because the clumpy graphite ring 
blocks the direct view of the central nucleus along the line of sight.  
The best fit parameters for this model are $R_{\rm in,r}$ = 0.04 pc, $\tau_{9.7, \rm r}$= 10, 
$R_{\rm clump}$=0.002 pc, 
and $\sigma$=30$^\circ $ with $\chi^2_{\rm min}$= 1.277
(see Table \ref{A3} in Appendix A). 
This clumpy nature of the dust in the ring reduces the scattered light contribution to the UV/optical SED. 
As a result, there is deficiency of model SED flux for $\lambda$ $\leqslant$  1 $\mu$m as seen in Fig. \ref{radius_rat}, thereby resembling an
obscured AGN SED.
For $\lambda > $ 1 $\mu$m, the clumpy RAT model gives a reasonable fit to the observed SED of NGC 4151.
     


\subsubsection{Smooth RAT model}
\begin{figure}
\centering
	\includegraphics[width=\columnwidth]{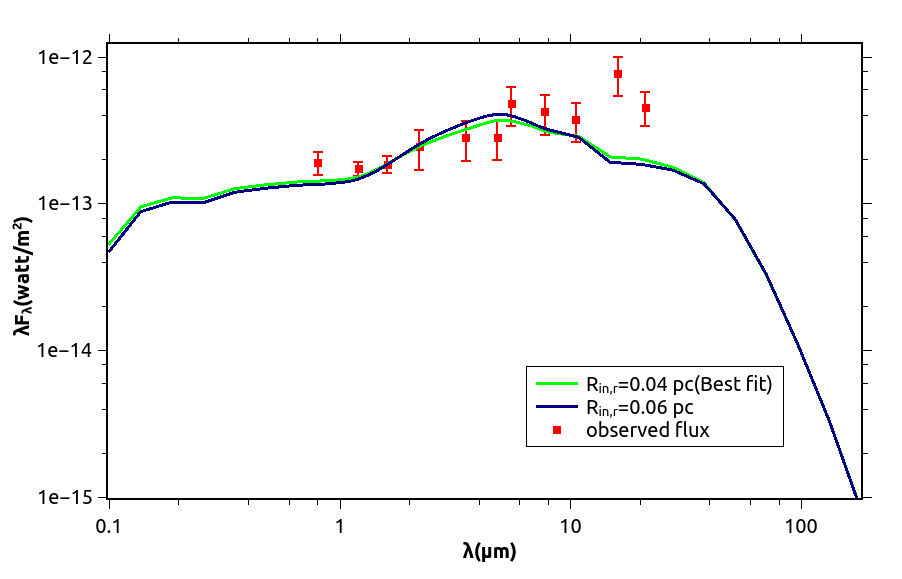}
\caption{The results from smooth RAT model up to 21 $\mu$m for two different ring inner radii. The best fit corresponds to the
	ring radius of 0.04 pc. }
	\label{smoothring}
\end{figure}

\begin{figure}
\centering
\includegraphics[width=\columnwidth]{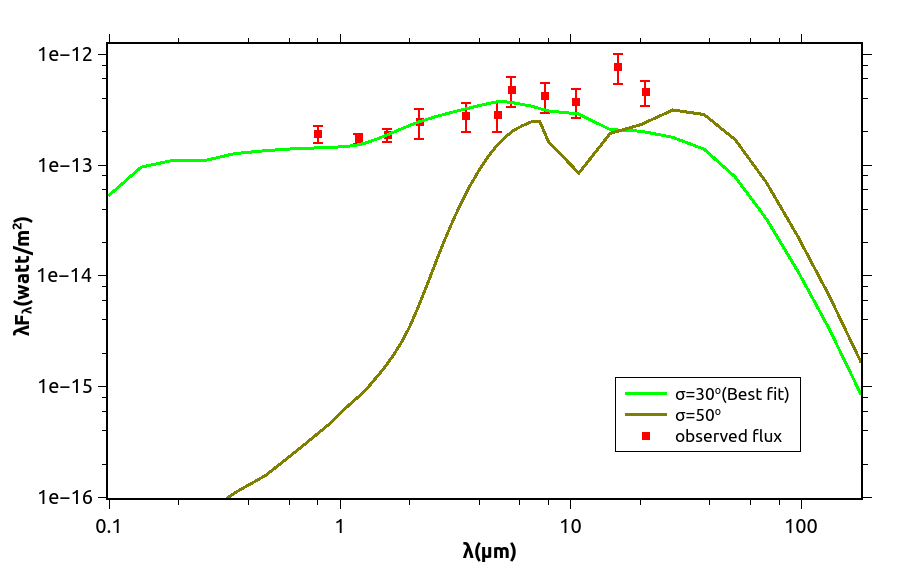}
\caption{The change in SED of smooth RAT model with varying half opening angle of the torus. The rest of the parameters are
	the best fit parameters corresponding to the $\chi^2_{\rm min}$.
}
\label{halfop}
\end{figure}

In this model, a smooth distribution of graphite dust is considered in the ring.
Again, the inner radius of the ring is varied from 0.01 pc to 0.06 pc and the optical depth of the ring is varied from 0.1 to 100. 
Fig. \ref{smoothring} shows the model SEDs for different ring radii together with the observed SED.
The variation of the model SED with half-opening angle is shown in Fig. \ref{halfop}.
As the half opening angle increases from $30^{\circ}$ to $50^{\circ}$, the flux decreases in the NIR part with enhanced silicate absorption at 10$\mu m$. For 30$^{\circ}$ half opening angle of the torus, the observer's line of sight is not obstructed by the 
dust in the torus. This is because near the sublimation zone the height of the torus is comparable to that of the ring. 
So for $30^{\circ}$ half opening angle, we directly observe the NIR emitting region, but when the half opening angle of the torus is 50$^{\circ}$, there could be clumps inside the torus, which intercept the observer's line of sight. 
The orientation of the ring towards the observer is not known but we have considered the torus and the ring to be in the same plane and the outer radius of ring is not constrained to be in contact with the inner radius of the torus. 
The best fit inclination angle is found to be 52$^{\circ}$ corresponding to minimum $\chi^2$ value of 1.05
(see Fig. \ref{fig:figs2} and Table \ref{tab:tabs2}). This value of $i$ is within the error bar of the
corresponding value derived from TO model. Hence, $i$ is fixed at 53$^{\circ}$ in this model.

\begin{figure}
\centering
\includegraphics[width=\columnwidth]{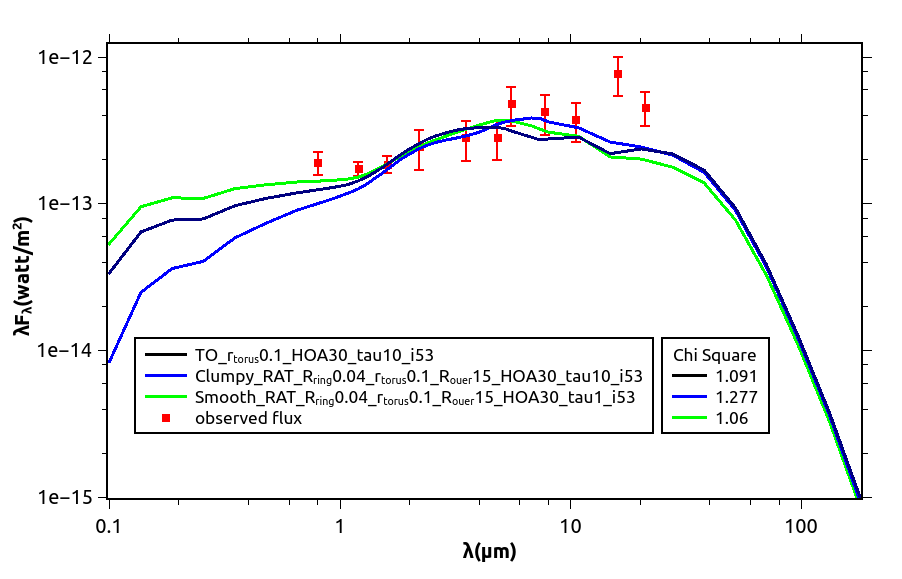}
\caption{This figure shows the comparison between the SEDs of the TO, the clumpy RAT and 
	the smooth RAT models. The best fit parameters are 
$R_{\rm in,\rm t}$=0.1 pc, $\tau_{9.7, \rm t}$ = 10 for the TO model,
$R_{\rm in,r}$ = 0.04 pc, $\tau_{9.7, \rm r}$= 10 for clumpy RAT model and 
$R_{\rm in,r}$ = 0.04 pc, $\tau_{9.7, \rm r}$= 1 for smooth RAT model.
}
\label{comp_all}
\end{figure}

The best fit parameters from the reduced $\chi^2$ test for the ring are $R_{\rm in, r}$ = 0.04 pc, $\sigma$ = 30$^{\circ}$, $\tau_{9.7,\rm r}$ = 1 (see Table \ref{A4} in Appendix A). The total optical depth in this model is 11 where it is 10 for torus and 1 for the ring. The best fit SEDs for all the models and the observed SED are 
plotted in Fig. \ref{comp_all}.
 The model value obtained for the ring radius $\sim$ 0.04 pc is in good agreement with the observed innermost radius of the torus $\sim$ 0.04 pc from \cite{swain2003interferometer} and \cite{suganuma2006reverberation}. 
The observed and the model derived inner radius of NGC 4151 torus, are listed in the Table \ref{sub_tab} and Table \ref{DP} respectively.

\begin{figure}
\centering
\includegraphics[width=\columnwidth]{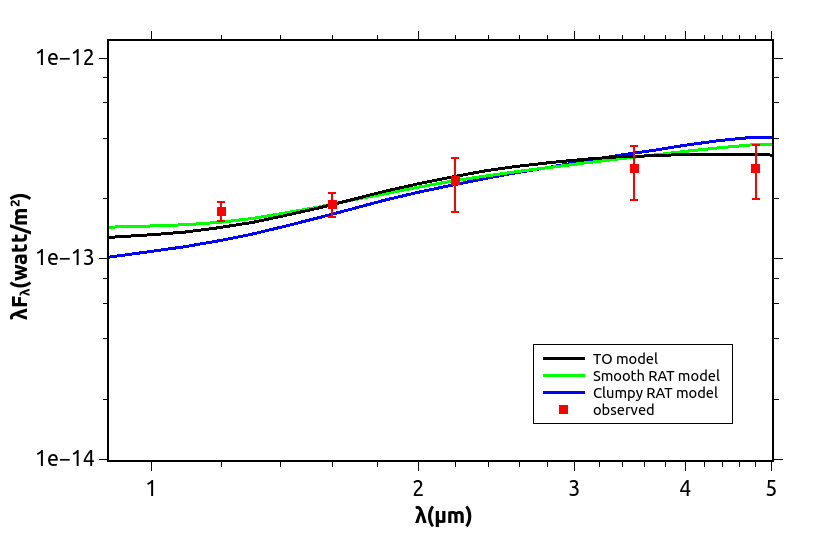}
\caption{The comparison of SEDs from TO, clumpy RAT and smooth RAT models with observed SED in NIR range.  
}
\label{compbrst}
\end{figure}
It is clear that the NIR model flux is a good fit to the observed flux as in the case of the earlier two models although the fit worsens in the MIR. 
This is evident in Fig. \ref{comp_all}, where the three model SEDs are plotted for comparison. We see that none of the
models account for the MIR emission for $\lambda $> 10$\mu $m. This is probably because other possible sources of MIR emission e.g., polar dust
\citep{2017ApJ...838L..20H}, which is known to exist in several AGN, haven't been accounted for, in the current work.  

From the above results it is evident that the parameters derived from the RAT models are not only physically 
acceptable but also in agreement with the observed properties of NGC 4151. On the other hand, the TO model, despite 
giving a reasonable $\chi^2 $ fit to the observed SED, results in an unphysical value of inner radius for a mixture of silicates and graphites.

A closer inspection of the NIR part of the SED is shown in Fig. \ref{compbrst}. Based on the minimum 
$\chi^2$ criterion, the smooth RAT model appears to be the best fit. However, the $\chi^2$ of all the three 
models are reasonably close in magnitude and are in good agreement with the observed SED.
In the optical/NIR part of the SED, the accretion disk also contributes together with the emission
and scattering from hot dust in the ring/torus. Due to the compactness of the region close to the accretion disk, it is not known whether the NIR emission is due to the accretion disk itself
or from hot dust close to the accretion disk or both. If hot dust is the source of the NIR emission, it is not clear whether it is a part of the torus, or
lies outside the BLR, or between the BLR and the torus. 
It can be noticed from Figs. \ref{fig:figs4}, 
\ref{fig:figs5} that the hot dust mainly contributes in the 2-5 $\mu $m region while the accretion disk starts dominating for wavelengths $<$ 2 $\mu $m. 

\begin{table*}
\begin{center}
\caption[]{The list of observed dust inner radius for NGC 4151. For reference 1: \cite{Gandhi:2015vta}, 2:\cite{minezaki2003inner}, 3:\cite{0004-637X-698-2-1767}, 4:\cite{Koshida_2014}, 5:\cite{refId1}, 6:\cite{Kishimoto_2013}} 
\begin{tabular}{|c|c|c|c}
\hline 
$R_{\rm in}$ (pc) & method & Reference \\
\hline 
 0.030$^{+0.008}_{-0.006}$ & Reverberation mapping &[1] \\
\hline 
0.041$\pm$ 0.004         & NIR interferometric techniques   & [1] \\
\hline         
$\sim$ 0.04               & Thermal dust reverberation        & [2]  \\
\hline  
< 0.1                  & Gemini NIR integral Field        & [3]  \\
                          & spectrograph observations                    &    \\
\hline
$\sim$ 0.04               & Dust reverberation mapping        & [4]  \\
\hline
$\sim$ 0.03               & NIR reverberation mapping         & [5]\\
\hline
$\sim$ 0.06               & Keck Interferometry         & [6] 
\label{sub_tab}
\end{tabular} 
\end{center}
\end{table*} 

\begin{table*}
\begin{center}
\caption[]{The list of model derived inner radius for NGC 4151. 1:\cite{selfcon}, 2:\cite{gonzalez2019exploring}, 3:\cite{doi:10.1111/j.1365-2966.2006.09866.x}, 4:\cite{2008ApJ...685..147N}.}
\label{DP}
\begin{tabular}{clcl}
  \hline\noalign{\smallskip}
 $R_{\rm in}$ (pc)       & Total $\tau_{9.7}$ & method & Reference \\      \\
  \hline\noalign{\smallskip}

$\sim$ 0.1       &   -    & Two phase model &     [1]      \\
\hline
< 3   &  & Two phase model  & [2] \\
     &    > 9.8      &   Smooth model &   [3]     \\
     &    < 10.2     &   Clumpy model  &  [4]     \\
\hline
0.04   & 11 & smooth RAT model & This work \\
0.04   & 20 & clumpy RAT model & This work \\
  \noalign{\smallskip}\hline

\end{tabular}
\end{center}
\end{table*}

\begin{table*}
\begin{center}
\caption[]{The list of derived parameters for NGC 4151 from the literature.  For reference , 1: \cite{2009ApJ...702.1127R}, 2:\cite{Ramos_Almeida_2011}, 3: \cite{selfcon} and 4:\cite{gonzalez2019exploring}. Column 1 shows the inclination angle $i$, column 2 represents the half opening angle $\sigma$, column 3 and 4 represents the dust distribution parameters (p, q), column 5 and 6 are the optical depth of the dust ($\tau_{9.7}$,$\tau_v$) which are normalized at visible and 9.7 $\mu$m respectively. The column 7 is about other parameters which are not common to all the model, where $\gamma$ is radial extent of the torus, $A^{LOS}_v$ is the optical extinction at line of sight, $N_{0}$ is the number of clouds along the equatorial ray, $\eta$ is filling factor for the torus, $\tau_{v,cl}$ is the cloud optical depth, $\tau_{v,mid}$ is the optical depth of the disk mid-plane, a is the size of grain, $a_w$,$h$,$f_w$ are the wind parameters used in \cite{2017ApJ...838L..20H} which is further used by \cite{gonzalez2019exploring}}

\begin{tabular}{|c|c|c|c|c|c|c|c|c}
\hline
$i$(degrees)                 & $\sigma$(degrees)       & p                    & q & $\tau_{9.7}$ & $\tau_{v}$ &Other parameters & Reference \\ 
\hline
41$^{+23}_{-28}$ & 32              & 1.7$^{+0.8}_{-0.9}$ & - &-&120$^{+55}_{-48}$ & $A_v^{LOS}$<210 & [1] \\
                 &                 &                           &  &                   & & $N_{0}$= <3      &       \\
                 &                 &                           &  &                   &                  & \\
\hline    
43$^{+18}_{-26}$ & 24$^{+17}_{-6}$ & 1.8$\pm$0.6               & - &- & 110$^{+23}_{-26}$ &$A_v^{LOS}$<60    &[2]\\
                 &                 &                           &  &                    & &$\gamma$=16$^{+18}_{-7}$ & \\ 
                 &                 &                           & &                    & &$N_{0}$=2$^{+2}_{-1}$ & \\
\hline

43$^{+19}_{-24}$ &        -          &                    &- &- & 40$^{+5}_{-39}$ ($\tau_{v,cl}$) &$R_{\rm in}$=0.1$^{+0.41}_{-0.02}$ & [3]\\
                 &                  &                    & & &                                       &$\eta$=3$^{+75}_{-2}$ \\
                 &                  &                    & & &                                       &$\tau_{v,mid}$= 230$^{+770}_{-190}$ \\
                 &                  &                    & & &                                     &Intrinsic $L_{AGN}$(LogL$\odot$)= \\ 
                 &                  &                    & & &                                                 &10.35$^{+0.02}_{-0.02}$ \\

\hline
10.2$^{+10.4}_{-7.1}$ & 20.1 &>-0 & - & >9.8 &- &$\Gamma$>6& [4] \\
                       &     &      &   &      &  & $\gamma$<10 & \\
\hline
>87.5                & 28.7$^{+35.6}_{-25}$& 1.64$^{+1.75}_{-1.58}$ & & <10.2 &  &$N_{0}$>14.1 & [4] \\
             &                 &                        & & &       &$\gamma$=10.1$^{+10.7}_{-9.9}$\\                                                                                 

\hline
69.6 $^{+78.4}_{-64.8}$ & >54.4 &-0.77$^{+0.73}_{-0.82}$ &- &-&36.3$^{+41.1}_{-33.8}$($\tau_{cl}$) &$N_{0}$=8.3$^{+9.5}_{-6.0}$& [4]\\
\hline                                                           
>86.0 &- &- &- &- & <0($\tau_{cl}$) & $\tau_{disk}$=13.5$^{13.51}_{13.5}$ & [4] \\
      & & & & &                   & $R_{in}$<3 pc                       &\\
      & & & & &                   & $\eta$=2.38$^{+2.41}_{-2.34}$    & \\

\hline
<16.1 & >79.6 & >1.5 &< 0  & >3.81$^{4}_{3.76}$ &-&  $\gamma$<10 & [4] \\

\hline 
59.9$^{60.8}_{59.1}$ &>14.1& 2.5$^{-2.49}_{-2.52}$ & - &- &- & $a_w$=-0.92$^{-0.89}_{-0.96}$  &[4]\\
                     &       &  &  &  & & $h$=0.4$^{0.41}_{0.39}$ & \\
                     &                 &                       &            & & &$f_w$0.38$^{0.41}_{0.36}$& \\
                     &                 &                       &        & & &  $N_0$= 7$^{7.1}_{-6.9}$\\
                    &                 &                       &      & &       &$\theta$>44.8\\
\hline
53$^{3}_{-11}$ & 30  & 1 & 0 &  - & -&$\eta$=0.25  & This work \\
  &      &   &   &    & &$R_{\rm out}$=15 pc & \\
  &      &   &   &    & &Derived $R_{\rm in}$=0.04pc, \\
  &      &   &   &    & &Derived $\tau_{9.7,\rm total}$ $=$ 11 \& 20\\
  &      &   &   &    & &from smooth and clumpy RAT model respectively.\\
\hline

\label{para_allmodel}
\end{tabular} 
\end{center}
\end{table*} 

\cite{0004-637X-698-2-1767} attributed the NIR excess in AGN to thermal dust emission and also concluded that the temperature of the 
hot dust does not depend on the AGN luminosity. Our model favours the idea that the NIR excess could be due to large hot graphite 
dust in a smooth medium.  Regarding the location of hot dust from our study,
we see that the dust in the innermost regions exists at a radius greater than the sublimation radius of pure graphite grains.
This is in reasonable agreement with \cite{schunlletal2013} who found that the size of the NIR emitting region of 
NGC 4151 is 0.1 pc. 
The radius of the nuclear region containing hot dust in NGC 4151 was resolved by \cite{0004-637X-698-2-1767} upto 4 pc. 
The optical depth at V band from \cite{0004-637X-698-2-1767} is 1.86 which corresponds to $\tau_{9.7}=0.1$. Our derived best fit values for the optical depth, $\tau_{9.7, \rm r}$ are 1 and 10 for the smooth and the clumpy RAT models respectively. This 
difference in the values of the model and the observed optical depth has a negligible effect on the model SED. 

The half height of the torus is found to be less than 0.04 pc from Gemini Near-Infrared Integral Field Spectrograph (NIFS) observations \citep{0004-637X-698-2-1767}, which is close to the assumed height of the thin ring in the RAT model. 
For the best fit value of the inner radius of 0.04 pc, the temperature is $\sim$ 1295 K 
from the radius-luminosity relation. When the luminosity of NGC 4151 is increased to 10$^{44}$\,erg\, s$^{-1}$, as it is a variable AGN, the temperature of graphite becomes $\sim$ 1800K. The dust sublimation radius can be larger than the actual inner radius of sublimation zone observed by K-band reverberation mapping due to re-formation of the dust as reported by \cite{koshida2009variation}. There are many theoretical models for deriving the inner radius of AGN torus. The discrepancy between different inner radii derived theoretically solely depends on the chosen model. The SED depends on viewing angle for wavelengths shorter than FIR, due to considerable extinction at these wavelengths. Our results suggest that the graphite ring structure could be a separate entity or could be a part of the torus itself. 

The SED of our smooth/clumpy RAT model resembles a quasar SED where hot dust contributes to the NIR excess \citep{Zhuang_2018}. This means that the same scenario could also explain the dust composition in distant quasars.

The luminosity of AGN is fixed in our model. The increase in the luminosity only changes the scale of flux since the spectrum is the result of an interplay between luminosity and other parameters in the model.

Theoretically, the inner radius of dusty torus is estimated to be 0.1 pc from radius-luminosity formula while
\cite{refId1} found that the time-lag radii from near-IR reverberation mapping is 3 times smaller than 0.1 pc 
assuming graphite grains of size 0.05$\mu$m at 1500K.
Hence, Kishimoto et. al 2009 had predicted that only larger grains can survive at radii 3 times smaller than
the estimated 0.1 pc.
From the smooth and clumpy RAT models, the best fit radius of 0.04 pc agrees with 
the result of \cite{refId1} after incorporating large grains of size 0.1-1 $\mu$m.

It is also in agreement with model derived radius of \cite{selfcon} (see Table \ref{para_allmodel} for a comparison of different theoretical models for NGC4151).
As we know, the clumpy medium and smooth medium produce similar SEDs \citep{Feltre2012}. In the present model, 
the smooth medium provides a good fit to the observed NIR data rather than the clumpy medium. However, 
when dust is heated in a smooth medium, it can be destroyed or swept away unless shielded by a clumpy medium around it. Hence, the innermost regime contains a smooth ring shielded by the clumpy
dust clouds of the torus.
\begin{table*}
\begin{center}
\caption{Correlation coefficients for the three models for the entire wavelength range : Column 2 and 3 shows the best fit parameters. The symbols `$r$'  and `$p_{\rm r}$' represent Pearson coefficient and its probability respectively  where `$\rho$' and `$p_{\rm \rho}$' represent Spearman coefficient and its probability value respectively.} 
\begin{tabular}{|c|c|c|c|c|c|c|}
\hline 
Model & $R_{\rm in}$ & $\tau_{9.7}$ & $r$,$p_{\rm r}$ & $\rho$, $p_{\rm \rho}$ & $\chi^2$ value,$p$  \\
\hline 
TO & 0.1 pc & 10 & 0.2885, 0.3895 & 0.4181, 0.2031 & 1.003,0.4348 \\
\hline
Smooth RAT & 0.04 pc & 11 & 0.2007, 0.554 & 0.4909 , 0.1292 & 1.06,0.3882\\
\hline
Clumpy RAT & 0.04 pc & 20 & 0.628, 0.0386 & 0.6818 , 0.0255 & 1.277,0.2434 \\
\hline
\label{sub_tab2}
\end{tabular} 
\end{center}
\end{table*}

In summary, the smooth RAT model consists of a ring composed of large dust 
grains of pure graphite in smooth form and then a torus composed of mixture of graphite and
silicate grains in a two phase form both of which contribute to the output SED.
In general, all the models have similar $\chi^2$ values ($\sim$1). 
Hence, based on the $\chi^2$ criterion alone, it cannot be claimed that smooth RAT model 
is best model among all, as clumpy graphite dust has higher chance
 of existing in the inner regions of AGN.\ 
On the other hand, the SED obtained using $50^{\circ}$ half opening angle from Fig \ref{halfop} shows 
elevated MIR emission and a suppressed NIR emission.
This shows that in order to explain the observed MIR SED, we need to consider 
the polar dust, driven by radiation pressure which flow out as polar wind to emit in mid-IR 
around the accretion disk and which could also give rise to a high MIR emission even for low 
half opening angles.

\subsection{Statistical analysis}
\begin{figure*}
\centering
\includegraphics[width=5.5cm]{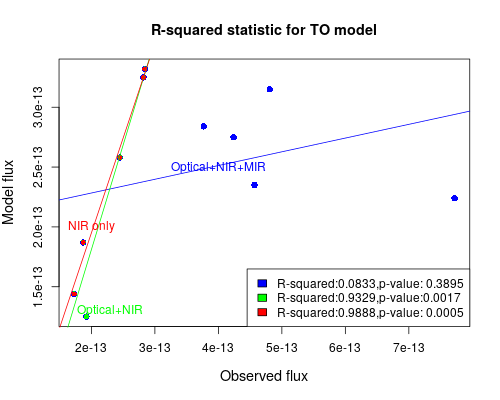}
\includegraphics[width=5.5cm]{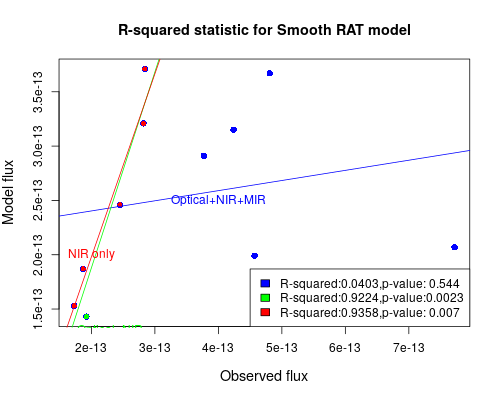}
\includegraphics[width=5.5cm]{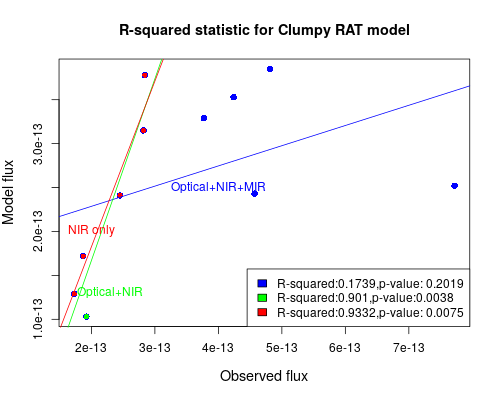}
\caption{
	The Figures of R$^2$ statistics for TO model ($R_{\rm in, \rm t}$ =0.1 pc, $\tau_{9.7, \rm t}$ =10), the smooth RAT model
($R_{\rm in, r}$ =0.04 pc, $\tau_{9.7,\rm r}$ =1) and the clumpy RAT model(($R_{\rm in, \rm r}$ =0.04 pc, $\tau_{9.7,\rm r}$ =10) 
	are presented. The blue line is fitted for the entire wavelength range (Opt+NIR+MIR), the green line for
	optical+NIR range and the red line is for NIR range only. The corresponding data for the fitted lines
	are represented by scatter plots of respective colors.}
\label{fig:rstat}
\end{figure*}

In order to assess how well our models represent the actual scenario in the innermost regions of NGC 4151, we have carried out the Pearson and Spearman correlation coefficient analysis. 
The values are listed in Tables \ref{sub_tab2}, \ref{sub_tab3} 
\& \ref{sub_tab4} for the best fit models derived using the reduced-$\chi^2$ test.
The Spearman and Pearson correlation coefficient tests were conducted to assess the level of linear relationship between the observed and the
estimated data values. 
Significantly positive correlation ($i.e.,$ $p < 0.05$) indicates a strong linear relationship between the two data sets.
We find that model and observed flux are significantly positively correlated only for the clumpy RAT 
model, when the entire wavelength range is considered (Table. \ref{sub_tab2}). Hence, we separated the optical+NIR, NIR and MIR data in order to check their individual contributions  
for all the models. For both optical+NIR
and NIR, we find that all the three models are positively correlated (Tables. \ref{sub_tab3} and 
\ref{sub_tab4}).
Further, linear relationship between the data values while accounting for variance was computed using regression analysis
 between the parameters $R_{in}$ and $\tau_{9.7}$ corresponding to each model (Fig.\ref{fig:rstat}).
The model flux ($f_{\rm mod}$) computed with different parameters and the observed flux ($f_{\rm obs}$) are fitted 
using the following equation :
\begin{equation}
	 f_{\rm mod} = a \times   f_{\rm obs}  + C,
\end{equation}
where $a$ is the slope and C is the intercept. 
From this analysis, we 
find that the `NIR only' and the `Optical+NIR' (in Fig.  \ref{fig:rstat}) are best fitted with the observed flux with $R^2 \sim 0.9$ and 
$ p < 0.05$ for all the three models. The best fit parameters are: TO model ($R_{\rm in,\rm t}$=0.1 pc and $\tau_{9.7,\rm t}$=10); smooth RAT model ($R_{\rm in,\rm r}$=0.04 pc and $\tau_{9.7,\rm r}$=1);
 clumpy RAT model ($R_{\rm in,\rm r}$=0.04 pc and $\tau_{9.7,\rm r}$=10) (see Appendix \ref{A5}). From this analysis, it can also be inferred that the
TO model gives the highest $R^2$ value of $\sim 0.9888$ and $p < 0.05$. But, the TO model is not physically feasible 
because of the destruction of ISM type dust close to the centre of the AGN. 
However, of the remaining two models, NIR flux in smooth RAT model accounts for 93.58\% ($p < 0.007$) of the observed flux whereas it is 93.38\%($p < 0.008$) in clumpy RAT model. 
In addition, from $\chi^2$ analysis although the smooth RAT model seems to be having the least $\chi^2$
that explains the observed data, in reality, considering both the $\chi^2$ and the $R^2$ tests, 
either of the models can explain the observed flux. 


\smallskip

\smallskip


\begin{table*}
\begin{center}
\caption{Correlation coefficients for the three models in Optical+NIR range(0.8-5 $\mu m$) : Column 2 and 3 shows the best fit parameters. The symbols `$r$'  and `$p_{\rm r}$' represent Pearson coefficient and its probability respectively  where `$\rho$' and `$p_{\rho}$' represent Spearman coefficient and its probability value respectively.}
\begin{tabular}{|c|c|c|c|c|c|}
\hline 
	Model & $R_{\rm in}$ & $\tau_{9.7}$ & $r$,$p_{\rm r}$ & $\rho$, $p_{\rm \rho}$   \\
\hline 
TO & 0.1 pc & 10 & 0.9659, 0.0017 & 0.8286, 0.0583  \\
\hline
Smooth RAT & 0.04 pc & 11 & 0.9604, 0.0023 &  0.8286, 0.0583 \\
\hline
Clumpy RAT & 0.04 pc & 20 & 0.9492, 0.0038 & 0.8286, 0.0583 \\
\hline
\label{sub_tab3}
\end{tabular} 
\end{center}
\end{table*} 

\begin{table*}
\begin{center}
\caption{Correlation coefficients for the three models in NIR range(1-5 $\mu m$) : Column 2 and 3 shows the best fit parameters. The symbols `$r$'  and `$p_{\rm r}$' represent Pearson coefficient and its probability respectively  where `$\rho$' and `$p_{\rm \rho}$' represent Spearman coefficient and its probability value respectively.} 
\begin{tabular}{|c|c|c|c|c|c|}
\hline 
	Model & $R_{\rm in}$ & $\tau_{9.7}$ & $r$,$p_{\rm r}$ & $\rho$, $p_{\rm \rho}$    \\
\hline 
TO & 0.1 pc & 10 & 0.9923, 0.0008 & 1, 0.0167  \\
\hline
Smooth RAT & 0.04 pc & 11 & 0.9674, 0.007 & 1, 0.0167  \\
\hline
Clumpy RAT & 0.04 pc & 20 & 0.9690, 0.0075 & 1, 0.0167  \\
\hline
\label{sub_tab4}
\end{tabular} 
\end{center}
\end{table*} 



\section{Conclusion}

In this paper, we have used $SKIRT$, a 3D radiative transfer code to model the dust emission of the NGC 4151 AGN.  
We have considered three models : the TO, the smooth RAT and the clumpy RAT models, all of which give considerably good fits to the observed SED
despite the different compositions and grain size distributions. We have successfully modelled the NIR excess of NGC 4151 using a mixture of graphite and
silicate grains of sizes in the range 0.005 - 0.25 $\mu$m for the TO model and an additional ring consisting of pure graphite dust of size range $\sim $
0.1-1 $\mu$m in the RAT models. Our main findings for NGC 4151 are as summarized below :.
\begin{itemize}
\item We find that the TO,
 the smooth RAT and the clumpy RAT models, all of them reproduce the features of the observed NIR SED 
		qualitatively, quantitatively and physically, based on the $\chi^2$ and the $R^2$ tests. 
\item The location of graphite dust is found to be 0.04 pc from the 
central source which is
in agreement with the observed radius for NGC 4151 according to both
smooth RAT model and clumpy RAT models from $\chi^2$ test. The smooth RAT
model explains the flat NIR emission better than the clumpy one which is further confirmed from R$^2$ test. 
	However, the clumpy RAT model also gives the location of graphite dust to be at 0.04 pc from $\chi^2$ test 
		and 0.06 pc from the $R^2$ test. 
\item From the smooth RAT model, the best fit parameters are found to be $i$=53$^\circ $ $^{3}_{-11}$, $R_{\rm in,\rm r}$=0.04 pc, 
$\sigma$=30$^\circ$ and $\tau_{9.7,\rm total}$=11. The $\chi^2$ values for all the models are reasonably close
and the smooth RAT model has least of them.
We conclude that both the RAT models can explain the observed SED as each model has its own parameter space. 
\item The best fit optical depth are found to be $\tau_{9.7,\rm r} = 1$ and 10 for the smooth and clumpy RAT models 
	respectively with $\tau_{9.7,\rm t} = 10$.
\item There is no significant difference in the values of $i$ for smooth and clumpy RAT models. The slight differences between the models under different dust morphology considerations, depends more on 
model parameters like inner radius of sublimation zone, optical depth of dust medium and on model 
assumptions such as chemical composition in each geometry rather than the type of distribution of dust (smooth or clumpy) in the ring.
\item The temperature of hot dust in the graphite ring is found to be $\sim$ 1295 K and could be as high as $\sim$ 1800 K for source luminosity of 10$^{44}$ ergs\ s$^{-1}$.
\item In our model, the orientation of the ring is assumed to be co-planar with the torus axis. The hot graphite dust may or may 
	not be a part of the torus or could be oriented in a different geometry which can play a significant role in fitting the
		observed SED.
\item  In the TO model, $R_{\rm clump}$ is not a sensitive parameter, however, it is a sensitive parameter in the clumpy RAT model.
\item The best fit half-opening angle for the torus is $\sigma = 30^{\circ}$, which agrees with the observations.
\item None of the models are able to explain the MIR part of the SED of NGC 4151. The MIR part of the SED could be better fitted if we incorporate
the polar dust winds in the model. This polar cone (\cite{2017ApJ...838L..20H}, \cite{stalevski2019dissecting}) could be 
		emanating from the graphite ring considered in our RAT model.

\end{itemize}
To conclude, this is the first time a detailed 3D modelling has been carried out for the innermost regions of NGC 4151 by SKIRT code and the existence of a ring structure has been shown to explain the observed NIR flux. There could be other contributions to the NIR flux like contribution from the jet or from star formation which is beyond the scope of the present work. In addition, the inclusion of polar dust winds contributing to the MIR emission can also improve the model further. 
\section{Acknowledgements}
 We thank the anonymous referee for his/her invaluable comments which helped us to improve this paper 
 significantly. We are grateful to Marko Stalevski for useful discussions and thoughtful comments. We thank Dr. Debbijoy Bhattacharya for helpful suggestions on the 
 manuscript. We also thank Prof. Prajval Shastri for her support.  We have made use of NASA/IPAC Extragalactic Database(NED) for our research work which 
 is operated by the Jet Propulsion Laboratory, California Institute of Technology. One of the authors (PS) acknowledges Manipal Centre for 
 Natural Sciences, MAHE for facilities and support.




\section{Data availability}
The compiled SED from the observed data underlying this article can be retrieved from  
\cite{2003AJ....126...81A}. The original observed data are
 available in SIMBAD via link \url{http://simbad.u-strasbg.fr/simbad/sim-ref?querymethod=bib&simbo=
 on&submit=submit+bibcode&bibcode=2003AJ....126...81A}.\\
 The model SED data 
will be shared on a reasonable request to the authors(\url{s.subhashree00@gmail.com}).

\bibliographystyle{mnras}
\bibliography{reference} 




\appendix

\section{}
The $\chi^{2}$, the Spearman/Pearson correlation tables for the three models are presented.
\begin{table*}
\centering
\begin{tabular}{|c|c|c|c|c|c|}
\multicolumn{5}{c}{TO model}\\
\hline 
 $\tau_{9.7}$ & \multicolumn{4}{c|}{HOA($\sigma$ in $^{\circ}$)} \\ \cline{2-5} 
 & 20 & 25 & 30 & 50  \\
\hline
10 & 1.274, 0.2451 & 1.053, 0.3944 & 1.003, 0.4348 & 26.38, 4.4962$\times$10$^{-50}$ \\ 
\hline 
\end{tabular}
\caption{$\chi^{2}$ table for $\tau_{9.7}$=10, $R_{\rm in}$ = 0.06 pc, $R_{\rm out}$ = 15 pc, $i$= 53$^{\circ}$, $R_{\rm clump}$  = 0.4 pc and $p$ = 1 and $q$ = 0.   with varying half opening angle($\sigma$). For each entry, the first value is the $\chi^2$ and second value represents the probability. }
\label{A1} 
\end{table*}

\begin{table*}
\centering
\begin{tabular}{c|c|c|c}
\multicolumn{4}{c}{TO model}\\
\hline
$\tau_{9.7}$ & \multicolumn{3}{c}{Varying inner radius of the torus($R_{\rm in}$ in pc)} \\ \cline{2-4}
    & 0.04  & 0.06  & 0.1 \\ 
\hline 
\hline 
0.1 &5.615, 8.54$\times$10$^{-8}$ & 4.941, 1.16$\times$10$^{-7}$ & 3.43, 0.0003\\	
1 & 2.756, 0.0032 &	2.834, 0.0024 &	2.66, 0.0044 \\
5 & 1.415, 0.1749 & 1.313, 0.2238 & 1.280, 0.2417 \\
10 &1.006, 0.4323 &	1.003, 0.4348 & 1.09, 0.3661\\
15 & 1.365, 0.1977 & 1.262, 0.252 & 1.152, 0.3225\\
\hline
\end{tabular}
\caption{$\chi^2$ table for $\sigma$ = 30$^{\circ}$ with varying $R_{\rm in}$ and optical depth of the torus. For each entry, the first value is the $\chi^2$ and second 
	value represents the probability.}
\label{A2} 
\end{table*}

\begin{table*}
\centering
\begin{tabular}{|c|c|c|c|c|}
\multicolumn{5}{c}{Clumpy RAT model}\\
\hline 
$\tau_{9.7,\rm r}$  & \multicolumn{4}{c|}{Varying inner radius of the ring($R_{\rm in}$ in pc)} \\ \cline{2-5}		
& 0.03 & 0.04 & 0.05 & 0.06\\
\hline 
\hline 
0.1 & 1.496, 0.1427 &   1.412, 0.1763 & 1.492, 0.1442 &	1.571, 0.1175\\
1   & 1.344, 0.2079 &	1.481, 0.1783 & 1.597, 0.1097 &	1.712, 0.0803\\
10  & 1.609, 0.1062 &	1.277, 0.2434 & 1.328, 0.2160 &	1.378, 0.1916\\
100 & 1.514, 0.1363 &	1.636, 0.0988 & 1.64, 0.0977  &	1.643, 0.0969\\
\hline
\end{tabular}
\caption{$\chi^2$ table for $\sigma$ = 30$^{\circ}$, torus optical depth = 10 with varying $R_{\rm in}$ of the ring and varying optical depth of ring. For
	each entry, the first value is the $\chi^2$ and second value represents the probability.} 
\label{A3}
\end{table*}
\begin{table*}
\centering
\begin{tabular}{|c|c|c|c|c|}
\multicolumn{5}{c}{Smooth RAT model}\\
\hline 
$\tau_{9.7,\rm r}$  & \multicolumn{4}{c|}{Varying inner radius of the ring($R_{\rm in}$ in pc)} \\ \cline{2-5}		
& 0.03 & 0.04 & 0.05 & 0.06\\
\hline 
\hline 
0.1 & 1.356, 0.2021 &	1.203, 0.2878 &1.293, 0.2345 &	1.383, 0.1893 \\
1 & 1.329, 0.2155	&1.061, 0.3882 & 1.197, 0.2916 &	1.333, 0.2135 \\
10 & 1.108, 0.3528 &1.181, 0.3019	& 1.358, 0.2011 & 1.534, 0.1294 \\
100 &8.655, 4.23$\times$10$^{-13}$ &12.127, 2.28$\times$10$^{-19}$ &  10.496, 2.06$\times$10$^{-16}$ &	8.866, 1.78$\times$10$^{-13}$\\

\hline
\end{tabular}
\caption{$\chi^2$ table for $\sigma$ = 30$^{\circ}$, torus optical depth = 10 with varying $R_{\rm in}$ of the ring and varying optical depth of ring. For
	each entry, the first value is the $\chi^2$ value and second value represents the probability.} 
\label{A4}
\end{table*}

\begin{table*}
\centering
\begin{tabular}{|c|c|c|c|c|c|}
 \hline
\multicolumn{4}{c}{TO model}\\
\hline
$\tau_{9.7}$ & \multicolumn{3}{c}{Varying inner radius of the torus($R_{\rm in}$ in pc)} \\ \cline{2-4}
   & 0.04 & 0.06 & 0.1 \\ 
 \hline

 0.1 & 0.8677, 0.0213 & 0.8904, 0.01594 & 0.8683, 0.02115 \\
 \hline
 1 &  0.888, 0.01648 & 0.7991, 0.04079 &	0.8689, 0.02099 \\
 \hline
10 & 0.9762, 0.001566 &	0.9881, 0.000549	& 0.9888, 0.0005075 \\
 \hline
 15 &  0.982, 0.001028 &	0.9873, 0.0006077 & 0.9846, 0.0008153 \\
 \hline
 \hline
   \multicolumn{4}{c|}{Smooth RAT model}\\
\hline
$\tau_{9.7}$ & \multicolumn{3}{c}{Varying inner radius of the ring(R$_{in}$ in pc)} \\ \cline{2-4}
   & 0.03 & 0.04 & 0.06 \\
   \hline
 0.1 & 0.957, 0.0038 & 0.9646, 0.0029	& 0.9507, 0.0047 \\
 \hline
 1 & 0.9408, 0.0062 & 0.9358, 0.007 &	0.9598, 0.0035\\
 \hline
  10 &0.9395, 0.0064 & 0.9535, 0.0043 &0.9403, 0.0063\\
 \hline
 100 & 0.88, 0.0183 & 0.8337, 0.03035 & 0.8617, 0.02282\\
 \hline
 \hline
    \multicolumn{4}{c|}{Clumpy RAT model}\\
\hline
$\tau_{9.7}$ & \multicolumn{3}{c}{Varying inner radius of the ring(R$_{in}$ in pc)} \\ \cline{2-4}
  & 0.03 & 0.04 & 0.06 \\
  \hline
 0.1 & 0.931, 0.0079 & 0.9193, 0.009981 & 0.922, 0.0095\\
 \hline
 1 & 0.9297, 0.0081 & 0.9181, 0.0102	& 0.9158, 0.01064 \\
 \hline
  10 & 0.9247, 0.00897 & 0.9332, 0.0075 & 0.9385, 0.0066 \\
 \hline
 100 & 0.93, 0.0080 & 0.9245, 0.009	& 0.9248, 0.0089 \\
 \hline
 \end{tabular} 
	\caption{The R$^2$ table for TO model, smooth RAT model and clumpy RAT model.
	For each of the entries in the table, there are two values separated by comma, the first value being the coefficient of determination, R$^2$ and 
	the second value, the $p$-value} 
\label{A5}
\end{table*}
\clearpage

\newpage
\section{}

\begin{figure*}
\centering
\includegraphics[width=\columnwidth]{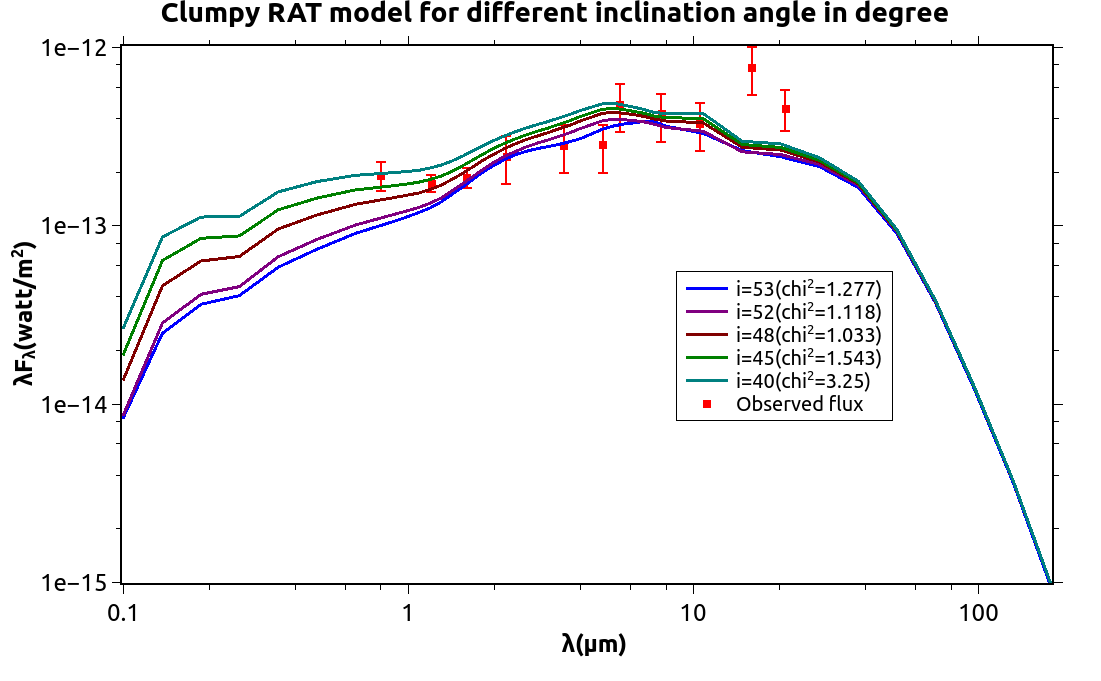}
\caption{The SEDs of clumpy RAT model for different
inclination angle.}
\label{fig:figs1}
\end{figure*}

\begin{figure}
\centering
\includegraphics[width=\columnwidth]{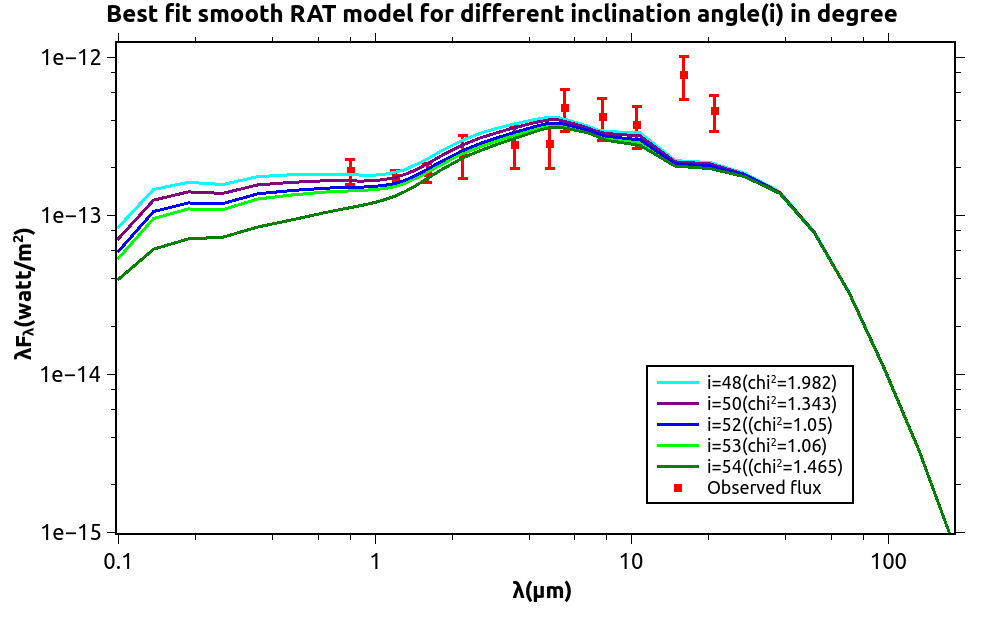}
\caption{The SEDs of smooth RAT model for different
inclination angle.}
\label{fig:figs2}
\end{figure}

\begin{figure}
\centering
\includegraphics[width=\columnwidth]{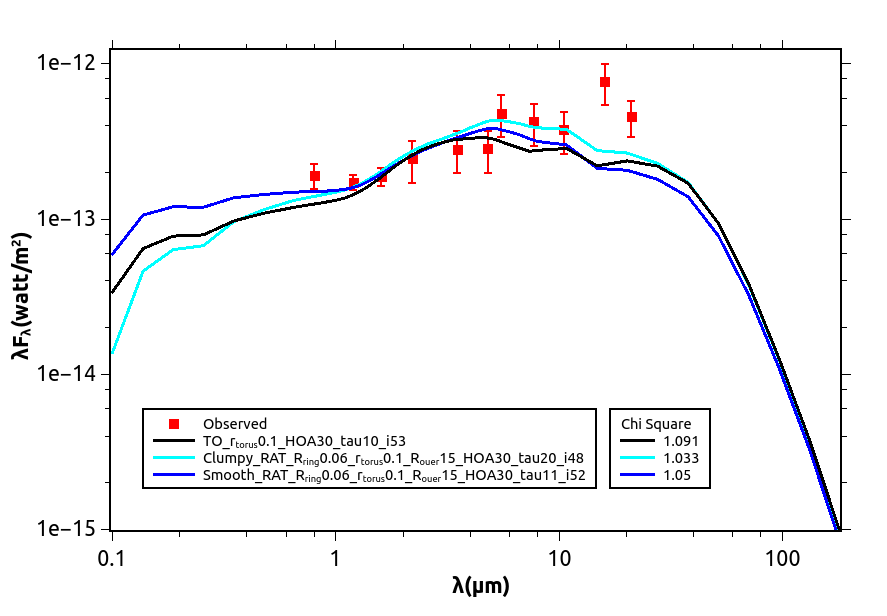}
\caption{The SEDs of all model for different
inclination angle.}
\label{fig:figs3}
\end{figure}

Here, We include the AGN model SEDs plot with varying inclination angle by keeping others parameters fixed in Fig. \ref{fig:figs1} and \ref{fig:figs2}. All the models with their best fit inclination angles are 
plotted in Fig. \ref{fig:figs3}. Although clumpy RAT model has the least $\chi^2$ value, smooth RAT model is well fitted with the observed flux in NIR band.\\

\begin{table*}
\centering
\begin{tabular}{|c|c|c|c|c|}
\multicolumn{5}{c}{Clumpy RAT model}\\
\hline 
$\tau_{9.7,\rm r}$  & \multicolumn{4}{c|}{Varying inner radius of the ring($R_{\rm in}$ in pc)} \\ \cline{2-5}		
& 0.03 & 0.04 & 0.05 & 0.06\\
\hline 
\hline 
0.1 & 1.058, 0.3905 &   1.143, 0.3278 & 1.145, 0.3264 &	1.146, 0.3257\\
1   & 1.076, 0.3767 &	1.161, 0.3154 & 1.463, 0.1552 &	1.129, 0.3376\\
10  & 1.033, 0.4103 &	1.032, 0.4111 & 1.089, 0.3668 &	1.145, 0.3264\\
100 & 1.063, 0.3867 &	0.92, 0.5062 & 1.046, 0.3999  &	1.049, 0.3976\\
\hline
\end{tabular}
\caption{$\chi^{2}$ table for i=48$^{\circ}$, $\sigma$ = 30$^{\circ}$, torus optical depth = 10 with varying $R_{\rm in}$ of the ring and varying optical depth of ring. For
	each entry, the first value is the $\chi^2$ and second value represents the probability.} 
\label{tab:tabs1}
\end{table*}

\begin{table*}
\centering
\begin{tabular}{|c|c|c|c|c|}
\multicolumn{5}{c}{Smooth RAT model}\\
\hline 
$\tau_{9.7,\rm r}$  & \multicolumn{4}{c|}{Varying inner radius of the ring($R_{\rm in}$ in pc)} \\ \cline{2-5}		
& 0.03 & 0.04 & 0.05 & 0.06\\
\hline 
\hline 
0.1 & 1.409, 0.1776  & 1.198, 0.2909 & 1.225, 0.2739 & 1.429, 0.1689 \\
1   & 1.403, 0.1802	& 1.051, 0.3960 & 1.174, 0.3066 & 1.355, 0.2025 \\
10  & 1.071, 0.3805 & 1.224, 0.2746	& 1.339, 0.2104 & 1.376, 0.1925 \\
100 & 8.264, 2.09$\times$10$^{-12}$ &11.705, 1.29$\times$10$^{-18}$ &  9.054, 5.36$\times$10$^{-21}$ &	8.474, 8.88$\times$10$^{-13}$\\

\hline
\end{tabular}
\caption{$\chi^{2}$ table for i=52$^{\circ}$, $\sigma$ = 30$^{\circ}$, torus optical depth = 10 with varying $R_{\rm in}$ of the ring and varying optical depth of ring. For
	each entry, the first value is the $\chi^2$ value and second value represents the probability.} 
\label{tab:tabs2}
\end{table*}

\begin{figure}
\centering
\includegraphics[width=7cm]{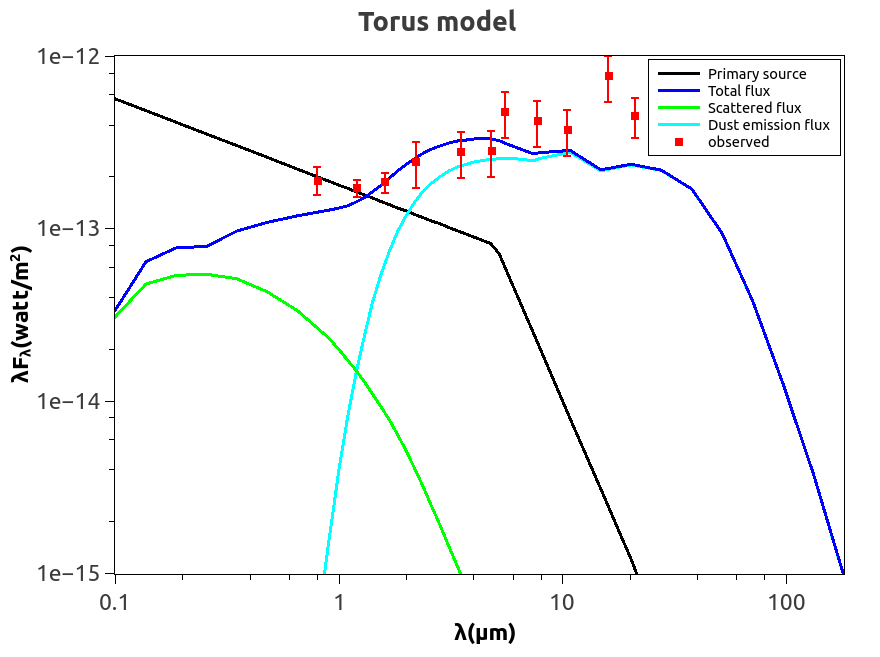}
\includegraphics[width=7.5cm]{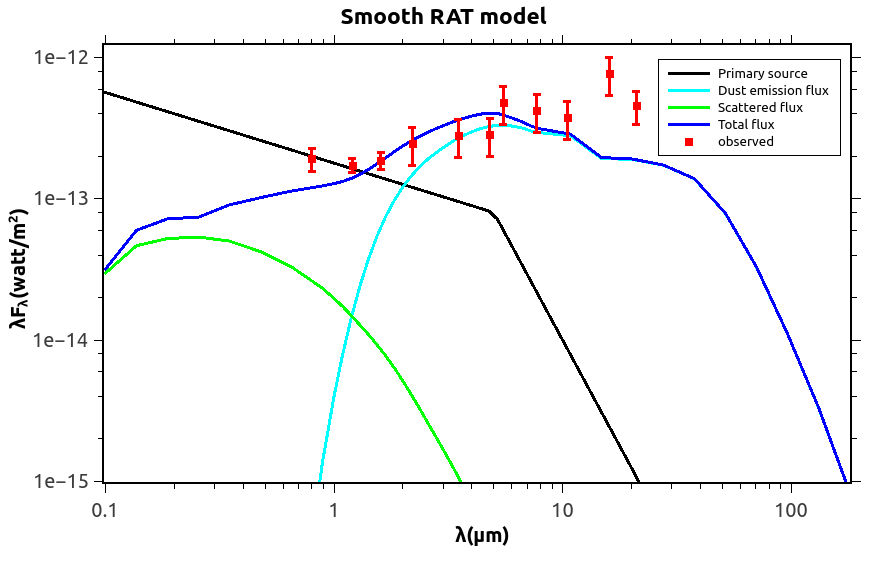}
\includegraphics[width=7cm]{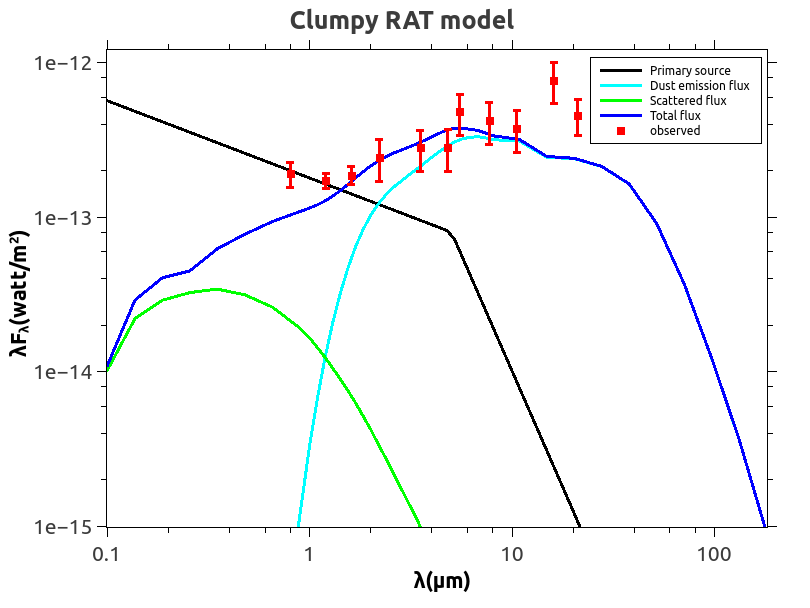}
\caption{The figure shows the contribution from the accretion disk, the ring and the torus to the SED for TO model, smooth RAT model and 
	clumpy RAT model respectively for the same parameters used in Figure \ref{comp_all}. The scale on X and Y axis is logarithmic.
}
\label{fig:figs4}
\end{figure}

\begin{figure}
\centering
\includegraphics[width=7cm]{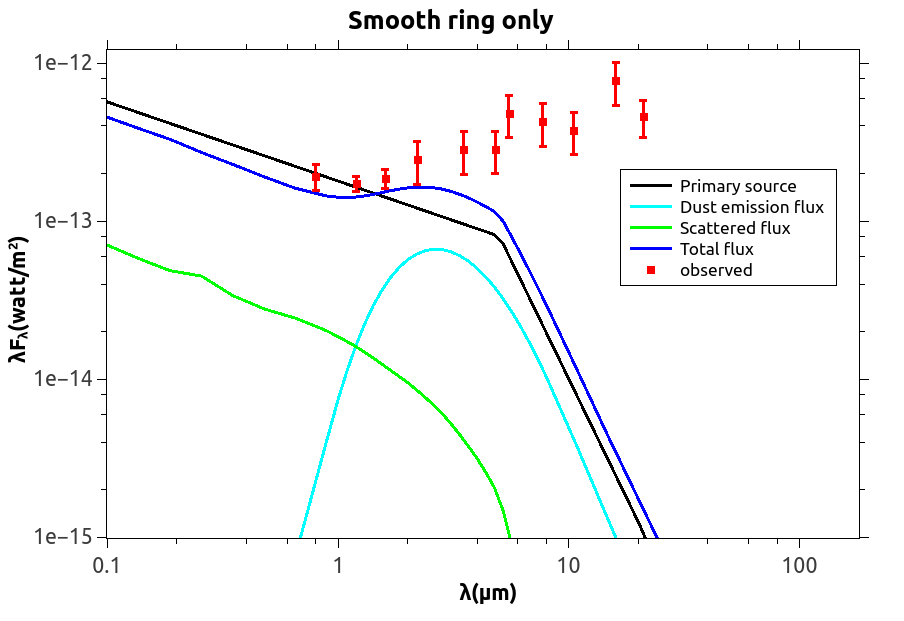}
\includegraphics[width=7cm]{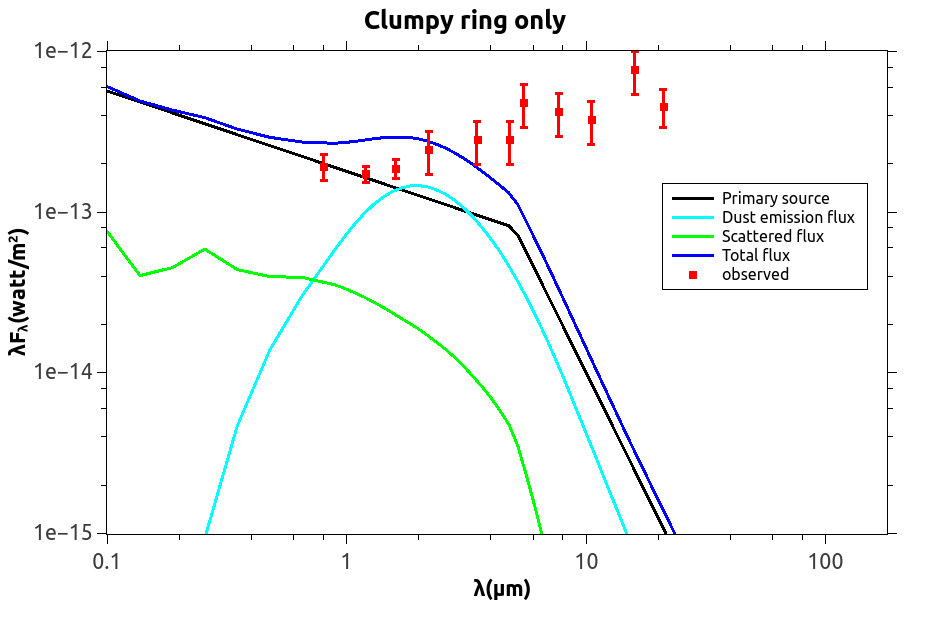}
\caption{The figure is showing the contribution from different components to the SED in smooth ring and clumpy ring model(which is without incorporating torus geometry) respectively for the parameters R$_{in,ring}$=0.04 pc, R$_{out}$=0.06 pc, tau=1, HOA=30$^{o}$. The scale on X and Y axis is in logarithmic scale.
}
\label{fig:figs5}
\end{figure}

\bsp	
\label{lastpage}
\end{document}